\documentclass{article}

\usepackage{amsfonts}

\def\diag{\mbox{diag}}

\def\mb{\mathbb}

\def\aa{\mathfrak{a}}
\def\bb{\mathfrak{b}}
\def\cc{\mathfrak{c}}
\def\dd{\mathfrak{d}}
\def\ee{\mathfrak{e}}
\def\ff{\mathfrak{f}}
\def\gg{\mathfrak{g}}
\def\hh{\mathfrak{h}}

\def\qa{\bf{a}}
\def\qb{\bf{b}}
\def\qc{\bf{c}}
\def\qd{\bf{d}}
\def\qe{\bf{e}}
\def\qf{\bf{f}}
\def\qg{\bf{g}}
\def\qh{\bf{h}}

\def\a{\hat{a}}
\def\b{\hat{b}}
\def\c{\hat{c}}
\def\d{\hat{d}}
\def\bea{\begin{eqnarray}}
\def\eea{\end{eqnarray}}
\def\({\left(}
\def\){\right)}
\def\<{\left<}
\def\>{\right>}

\def\tr{{\mbox{tr}}}
\def\bea{\begin{eqnarray*}}
\def\eea{\end{eqnarray*}}
\def\ben{\begin{eqnarray}}
\def\een{\end{eqnarray}}
\def\({\left(}
\def\){\right)}
\def\<{\left<}
\def\>{\right>}

\def\[{\left[}
\def\]{\right]}

\def\+{\bar}
\def\mb{\mathbb}
\def\tr{{\mbox{tr}}}

\def\L{{\cal{L}}}
\def\t{\widetilde}

\def\N{{\cal{N}}}
\def\M{{\cal{M}}}

\begin{document}
\setlength{\unitlength}{1mm}

\pagestyle{empty}
\vskip-10pt
\vskip-10pt
\hfill 
\begin{center}
\vskip 3truecm
{\Large \bf
Monopoles, three-algebras and ABJM theories\\ with $\N=5,6,8$ supersymmetry}\\ 
\vskip 2truecm
{\large \bf
Andreas Gustavsson\footnote{a.r.gustavsson@swipnet.se}}\\
\vskip 1truecm
{\it  Center for quantum spacetime (CQUeST), Sogang University, Seoul 121-742, Korea}\\
and\\
\it{School of Physics \& Astronomy, Seoul National University, Seoul 151-747 Korea}
\end{center}
\vskip 2truecm
{\abstract{We extend the hermitian three-algebra formulation of ABJM theory to include $U(1)$ factors. With attention payed to extra $U(1)$ factors, we refine the classification of $\N=6$ ABJM theories. We argue that essentially the only allowed gauge groups are $SU(N)\times SU(N)$, $U(N)\times U(M)$ and $Sp(N)\times U(1)$ and that we have only one independent Chern-Simons level in all these cases. Our argument is based on integrality of the $U(1)$ Chern-Simons levels and supersymmetry. A relation between monopole operators and Wilson lines in Chern-Simons theory suggests certain gauge representations of the monopole operators. From this we classify cases where we can not expect enhanced $\N=8$ supersymmetry. We also show that there are two equivalent formulations of $\N=5$ ABJM theories, based on hermitian three-algebra and quaternionic three-algebra respectively. We suggest properties of monopoles in $\N=5$ theories and show how these monopoles may enhance supersymmetry from $\N=5$ to $\N=6$.}}

\vfill 
\vskip4pt
\eject
\pagestyle{plain}

\section{Introduction}
There are evidences \cite{ABJM}, \cite{Kim:2009wb}, \cite{Kim:2010mr}  pointing to that the field theory dual to M-theory on $AdS_4 \times S^7$ is given by ABJM theory with gauge group $U(N)\times U(N)$ at Chern-Simons level $K=1$. However ABJM theory has only manifest $SU(4)\times U(1)$ R symmetry and $\N=6$ supersymmetry, as well as conformal symmetry \cite{Bandres:2008ry}. The dual field theory should have the same symmetries as the isometries of $AdS_4\times S^7$ for the duality to work. This means that we need to understand how the symmetries in ABJM theory can be enhanced to $SO(8)$ and $\N=8$ supersymmetry. 

From the field theory point of view, ABJM theory can be easily generalized to other gauge groups \cite{Schnabl:2008wj} while keeping the same amount of supersymmetry and R symmetry. In this Letter we choose to call these theories ABJM theories as well since the action for all these theories look exactly the same and only the gauge group is changed. One may then ask for which gauge groups and for which Chern-Simons levels there is enhanced supersymmetry.

 When the moduli space can be interpreted as M2 branes probing some orbifold singularity, one expects to have as R-symmetry the commutant of the discrete group of the orbifold singularity in $SO(8)$ \cite{Aharony:2008gk}. When the commutant is $SO(8)$ itself, one expects to have enhanced $\N=8$ supersymmetry. 

The reason for supersymmetry enhancement is the presence of monopole operators that transform in a suitable gauge representation \cite{ABJM}. In this paper we classify the gauge groups and Chern-Simons levels when suitable monopoles exist for enhancement from $\N=6$ to $\N=8$ supersymmetry. 

We extend the analysis to $\N=5$ ABJM theories. We find that these theories are described by the same type of Lagrangian as the $\N=6$ ABJM theories. The only difference is certain reality conditions on the matter fields. We show how suitable monopoles can enhance supersymmetry from $\N=5$ to $\N=6$. These theories with enhanced $\N=6$ supersymmetry should be dual to already known $\N=6$ ABJM theories \cite{Aharony:2008gk} but we will are unable to see to which $\N=6$ ABJM theory.  

After the first part of this paper was completed the paper \cite{Lambert:2010ji} appeared which also discusses $U(1)$-factors in the ABJM theory. Perhaps one goal was to establish that all ABJM theories can be formulated in terms of three-algebra. While this work could relate some of the $U(N)\times U(N)$ ABJM theories with $SU(N)\times SU(N)$ ABJM theories which do have a three-algebra formulation, it still misses out gauge groups $Sp(N)\times U(1)$ where a possible three-algebra formulation is left unclear. One purpose of this paper is to fill the gap and show that all ABJM theories have a three-algebra formulation. After the second part of this paper had been essentially completed, a paper \cite{Bagger:2010zq} appeared which also treats the $\N=5$ theories. More recently the paper \cite{Bashkirov:2010kz} appeared on supersymmetry enhancement to $\N=8$.

\section{ABJM theory and three-algebra}
ABJM theories can be formulated in terms of three-algebras \cite{BL} and in terms of Lie superalgebras \cite{Gaiotto:2008sd}, \cite{Hosomichi:2008jb}. For semi-simple Lie algebras the equivalence between the two has been established in \cite{deMedeiros:2008zh}, \cite{Palmkvist:2009qq}.

Bagger and Lambert \cite{BL} have given a three-algebra formulation of the supersymmetry variations for $\N=6$ supersymmetric ABJM theories with semi-simple gauge groups. The gauge group being semi-simple is very restrictive and the only candidate is $SU(N)\times SU(N)$. So we have to extend the three-algebra formulation so that it does not restrict ourselves only to semi-simple gauge groups. We will do this extension below. All  coupling constants are encoded in structure constants $f^{\bb \cc}{}_{\dd \aa}$ of the three-algebra. The action that was proposed does not apply to $U(1)$ factors in the gauge group. However the supersymmetry variations do. 

First we describe the ABJM Lagrangian as obtained by Bagger and Lambert. The Lagrangian is given by
\ben
\L &=& -D_{\mu} Z^A_{\aa} D^{\mu} Z_A^{\aa} + i\bar{\psi}^{A\aa} \gamma^{\mu}D_{\mu}\psi_{A\aa} - V(Z)\cr
&& + i\(\bar{\psi}^{A\aa} \psi_{A\bb} Z^B_{\cc} Z_B^{\dd} - 2 \bar{\psi}^{A\aa}\psi_{B\bb}Z^B_{\cc} Z_A^{\dd}\)f^{\bb\cc}{}_{\dd\aa}\cr
&& - \frac{i}{2}\(\epsilon^{ABCD} Z_A^{\aa} \bar{\psi}_{B\bb} \psi_{C\cc} Z_D^{\dd} + \epsilon_{ABCD} \bar{\psi}^{A\aa} Z^B_{\bb} Z^C_{\cc} \psi^{D\dd}\) f^{\bb\cc}{}_{\dd\aa}\cr
&& + \L_{CS}.\label{lag}
\een
Here 
\ben
\L_{CS} &=& \frac{1}{2} \(f^{\aa\bb}{}_{\cc\dd} A^{\cc}{}_{\bb} dA^{\dd}{}_{\aa} + \frac{2}{3} f^{\aa\cc}{}_{\dd\gg} f^{\gg\ee}{}_{\ff\bb} A^{\bb}{}_{\aa} A^{\dd}{}_{\cc} A^{\ff}{}_{\ee}\)\label{cs}
\een
is the Chern-Simons term and 
\bea
V(Z) &=& \frac{2}{3}\(f^{\aa\bb}{}_{\gg\hh} f^{\cc\hh}{}_{\ee\ff} - \frac{1}{2} f^{\aa\bb}{}_{\ee\hh} f^{\cc\hh}{}_{\gg\ff}\) Z^A_{\aa} Z_A^{\ee} Z^B_{\bb} Z_B^{\ff} Z^C_{\cc} Z_C^{\gg}
\eea
is the sextic potential. The covariant derivative is
\ben
D_{\mu}Z_{\aa} &=& \partial_{\mu}Z_{\aa} + A_{\mu}{}^{\dd}{}_{\cc} f^{\bb\cc}{}_{\dd\aa} Z_{\bb}. 
\een
In a convention where supersymmetry parameters $\epsilon_{AB}$ and $\epsilon^{AB}$ are such that
\ben
\epsilon^*_{AB} &=& -\epsilon^{AB},\cr
\epsilon^{AB} &=& \frac{1}{2}\epsilon^{ABCD} \epsilon_{CD}
\een
the $\N=6$ supersymmetry variations read
\ben
\delta Z^A_{\aa} &=& -i \bar{\epsilon}^{AB} \psi_{B\aa},\cr
\delta \psi_{A\aa} &=& \gamma^{\mu} \epsilon_{AB} D_{\mu} Z^B_{\aa}\cr 
&& - \(\epsilon_{AB} Z^B_{\bb} Z^C_{\cc} Z_C^{\dd} + \epsilon_{BC} Z^B_{\bb} Z^C_{\cc} Z_A^{\dd}\) f^{\bb\cc}{}_{\dd\aa},\cr
\delta \bar{A}_{\mu}{}^{\bb}{}_{\aa} &=& \(-i \bar{\epsilon}^{AB} \gamma_{\mu} \psi_{A\cc} Z_B^{\dd} + i\bar{\epsilon}_{AB} \gamma_{\mu} Z^A_{\cc} \psi^{B\dd} \) f^{\bb\cc}{}_{\dd\aa}.\label{N6}
\een

Infinitesimal gauge transformations read
\ben
\delta Z_{\aa} & = & \Lambda^{\dd}{}_{\cc} f^{\bb\cc}{}_{\dd\aa} Z_{\bb},\cr
\delta A_{\mu}{}^{\dd}{}_{\cc} & = & -D_{\mu} \Lambda^{\dd}{}_{\cc} 
\een
The only thing that we can adjust are the structure constants $f^{\bb\cc}{}_{\dd\aa}$. Supersymmetry alone does not fix the overall constant of the Lagrangian and so we could also imagine a rescaling of the Lagrangian by an overall factor $\lambda$,
\bea
\L &\rightarrow & \lambda \L
\eea
However such a rescaling is equivalent to instead scaling the Chern-Simons levels according to\footnote{Schematically the Lagrangian and covariant derivative are given by
\bea
\L &=& \lambda\Big(-(DZ)^2 + i \psi D \psi - f^2 Z^6 + f \psi^2 Z^2 + f AdA + f^2 A^3\Big),\cr
DZ &=& dZ + f A Z
\eea
If we replace $Z$ by $Z/\sqrt{\lambda}$, $\psi$ by $\psi/\sqrt{\lambda}$ and $A$ by $A/\lambda$, we get
\bea
\L &=& -(DZ)^2 + i \psi D \psi - \(\frac{f}{\lambda}\)^2 Z^6 + \frac{f}{\lambda} \psi^2 Z^2 + \frac{f}{\lambda} AdA + \(\frac{f}{\lambda}\)^2 A^3,\cr
DZ &=& dZ + \frac{f}{\lambda} A Z.
\eea
An overall rescaling of the action amounts to a rescaling of $f$. To see the relation to the Chern-Simons level we replace $fA$ with $A$,
\bea
\L &=& -(DZ)^2 + i \psi D \psi - f^2 Z^6 + f^2 \psi^2 Z^2 + \frac{1}{f}\(AdA+A^3\),\cr
DZ &=& dZ + A Z.
\eea
We conclude that the Chern-Simons level is $k \sim \frac{1}{f}$.}
\bea
k_l &\rightarrow & \lambda k_l
\eea
where $l$ runs over all the gauge group factors for each of which we a priori could have a different Chern-Simons level. As we will see, supersymmetry restricts these levels. For all $\N=5,6$ supersymmetric Chern-Simons theories it turns out there is exactly one independent Chern-Simons level (if one does not count decoupled direct sums of these theories).

The Lagrangian has $\N=6$ supersymmetry if and only if the three-algebra structure constants satisfy the fundamental identity as well as
\ben
f^{\bb\cc}{}_{\dd\aa} &=& -f^{\cc\bb}{}_{\dd\aa},\label{constraint}\\
f^{*\bb\cc}{}_{\dd\aa} &=& f^{\dd\aa}{}_{\bb\cc}.
\een
The fundamental identity can be expressed entirely in terms of the $f^{\bb\cc}{}_{\dd\aa}$ but is more transparent written in three-algebra notation. The three-algebra can be defined in an abstract way as 
\bea
[T^{\aa},T^{\bb};T^{\cc}] &=& f^{\aa\bb}{}_{\cc\dd} T^{\dd}
\eea
where the three-bracket is subject to the fundamental identity
\bea
&&[[T^{\aa},T^{\bb};T^{\cc}],T^{\ee};T^{\ff}]-[[T^{\aa},T^{\ee};T^{\ff}],T^{\bb};T^{\cc}]\cr
 &=& [T^{\aa},[T^{\bb},T^{\ee};T^{\ff}];T^{\cc}]-[T^{\aa},T^{\bb};[T^{\cc},T^{\ff};T^{\ee}]].
\eea
We have a matrix realization where the three-algebra generators are representented by matrices
\bea
(T^{\aa})_a{}^{\hat{a}}.
\eea
Their complex conjugates are denoted 
\bea
(T_{\aa})_{\hat{a}}{}^a.
\eea
The three-bracket is realized as a certain matrix multiplication
\bea
[T^{\aa},T^{\bb};T^{\cc}]_a{}^{\hat{a}} &\propto& (T^{\aa} T_{\cc} T^{\bb} - T^{\bb} T_{\cc} T^{\aa})_a{}^{\hat{a}}
\eea
One may check that this solves the fundamental identity. 

To the three-algebra is associated a Lie algebra, that corresponds to the gauge group of ABJM theory. The ABJM matter fields are in the bifundamental representation of the gauge group 
\bea
(Z^A)_{a}{}^{\hat{a}} &=& Z^A_{\aa} (T^{\aa})_a{}^{\hat{\a}} 
\eea
and the gauge field in the adjoint representation of the gauge group
\bea
A_{\hat{a}}{}^{\hat{b}} &=& A^{\dd}{}_{\cc} (T_{\dd})_{\hat{a}}{}^{a} (T^{\cc})_a{}^{\hat{b}}\cr
A_a{}^b &=& A^{\dd}{}_{\cc} (T^{\cc})_a{}^{\hat{a}} (T_{\dd})_{\hat{a}}{}^b.
\eea

\subsection{Abelian factors in three-algebra}
ABJM theory with gauge group $U(1)\times U(1)$ should have vanishing three-algebra structure constants as there is no scalar potential. Yet there is a gauge covariant derivative that involves the gauge potential. Since the derivative in three-algebra language was given as
\bea
D_{\mu} Z_{\aa} &=& \partial_{\mu} Z_{\aa} + A_{\mu}{}^{\dd}{}_{\cc} f^{\bb\cc}{}_{\dd\aa} Z_{\bb}
\eea
and this reduces to 
\bea
D_{\mu} Z_{\aa} &=& \partial_{\mu} Z_{\aa}
\eea 
if $f^{\bb\cc}{}_{\dd\aa} = 0$. The three-algebra formalution presented in \cite{BL} works only if the gauge group is $SU(N)\times SU(N)$. This is an unacceptable restriction that must be removed. If the restriction to semi-simple gauge groups is not removed, it makes three-algebra a useless tool for classifying ABJM theories. Fortunately the restriction can be removed and in fact it is quite easy to extend the three-algebra formulation so that it allows for $U(1)$ factors in the gauge group. This extended three-algebra formulation is then completely general and can be used to classify all ABJM theories. In the course of doing so, we refine and also correct a mistake in the previous classification of ABJM theories \cite{Schnabl:2008wj}. 
 
Let us define the gauge covariant derivative in terms of another gauge field $\bar{A}_{\mu}{}^{\bb}{}_{\aa}$ as
\ben
D_{\mu} Z_{\aa} &=& \partial_{\mu} Z_{\aa} + \bar{A}_{\mu}{}^{\bb}{}_{\aa} Z_{\bb}.\label{derivative}
\een
The gauge variation reads 
\ben
\delta Z_{\aa} &=& \bar{\Lambda}^{\bb}{}_{\aa} Z_{\bb},\cr
\delta \bar{A}_{\mu} &=& -D_{\mu} \bar{\Lambda}^{\bb}{}_{\aa}.\label{gaug}
\een
As far as closure of the supersymmetry variations concerns, we only need to consider the gauge field $\bar{A}_{\mu}{}^{\bb}{}_{\aa}$. Its supersymmetry variation was obtained in \cite{BL}. The difference to their formalism is that we drop the assumption that
\bea
\bar{A}_{\mu}{}^{\bb}{}_{\aa} &=& A_{\mu}{}^{\dd}{}_{\cc} f^{\bb\cc}{}_{\dd\aa}
\eea
As we saw above, assuming this leads to a contradiction if there are $U(1)$ factors. We only assume that this relation holds on the traceless parts. This is how we can extend the Bagger and Lambert three-algebra formalism to include $U(1)$ factors.

We decompose this gauge field into traceless and trace parts as
\ben
\bar{A}_{\mu}{}^{\bb}{}_{\aa} &=& \t B_{\mu}{}^{\bb}{}_{\aa} -i A^i_{\mu} e_i \delta^{\bb}_{\aa}.\label{tracedecomp}
\een
The derivative then becomes
\bea
D_{\mu} Z_{\aa} &=& \partial_{\mu} Z_{\aa} + \t B_{\mu}{}^{\bb}{}_{\aa} Z_{\bb} -i A^i_{\mu} e_i Z_{\aa}
\eea
and we identify $A^i$ as the gauge field in the $i$-th $U(1)$ factor, with the associated electric charge $e_i$. 

We also decompose the three-algebra structure constants into traceless and trace parts as
\ben
f^{\bb\cc}{}_{\dd\aa} &=& \t f^{\bb\cc}{}_{\dd\aa} + \lambda \delta^{\bb}_{\aa} \delta^{\cc}_{\dd}\label{lambda}
\een
where tracelessness means
\bea
\t f^{\bb\cc}{}_{\cc\aa} &=& 0,\cr
\t f^{\bb\cc}{}_{\dd\bb} &=& 0.
\eea

So far we have done nothing harmful to the Bagger and Lambert formulation. We have just decomposed their gauge field $\bar{A}_{\mu}{}^{\bb}{}_{\aa}$ and structure constants $f^{\bb\cc}{}_{\dd\aa}$ into traceless and trace parts. This means that we are still guaranteed to have on-shell closure in the Bagger and Lambert supersymmetry variations provided the total structure constants $f^{\bb\cc}{}_{\dd\aa}$ are antisymmetric and subject to the fundamental identity. In this sense we do not leave the three-algebra formalism, but we can also incorporate an arbitrary number of $U(1)$ factors into this formalism.

It is not true that the traceless structure constants $\t f^{\bb\cc}{}_{\dd\aa}$ are antisymmetric in $\bb$ and $\cc$. It is true though that these still satisfy the fundamental identity, and so does $\delta^{\bb}_{\aa} \delta^{\cc}_{\dd}$. 

We are now interested in the action that yields the equation of motion on which the $\N=6$ supersymmetry variations close. We make the following ansatz for the contribution from the $U(1)$ factors to the Chern-Simons part of the Lagrangian,
\ben
\sum_i \frac{k_i e_i^2}{4\pi} A^i dA^i \label{cs-ab}
\een
Here we find that we must take $k_i\in \mb{Z}$ for gauge group $U(1)$. We may integrate the Chern-Simons term on the manifold $S^1 \times S^2$. If there is a magnetic flux through $S^2$ of strenght
\bea
\int_{S^2} F^i &=& \frac{2\pi}{e^i}
\eea
we find that the Chern-Simons action becomes 
\bea
\sum_i k_i e_i \int_{S^1} A^i
\eea
and this is well-defined modulo $2\pi$ only if $k_i\in \mb{Z}$.

We can derive a constraint on the levels $k_i$ by varying the action and demand that the resulting equation of motion is what is needed in order to close supersymmetry on-shell. This equation of motion for the gauge field was obtained by Bagger and Lambert and it reads
\bea
\bar F_{\mu\nu}{}^{\bb}{}_{\aa} &=& -\epsilon_{\mu\nu\lambda} {J}^{\lambda}{}^{\dd}{}_{\cc} f^{\bb\cc}{}_{\dd\aa}
\eea
where
\bea
{J}^{\lambda}{}^{\dd}{}_{\cc} &=& Z^A_{\cc} D^{\lambda} Z_A^{\dd} - D^{\lambda} Z^A_{\cc} Z_A^{\dd} - i \bar{\psi}^{A\dd} \gamma^{\lambda} \psi_{A\cc}.
\eea
We are now interested in the trace part of this equation of motion. Using (\ref{tracedecomp}), (\ref{lambda}) we get
\ben
-i e_i F^i_{\mu\nu} &=& - \lambda \epsilon_{\mu\nu\lambda} {J}^{\lambda \cc}{}_{\cc}. \label{close}
\een
On the other hand, including the term (\ref{cs-ab}) into the Lagrangian, the equation of motion for the abelian gauge fields becomes
\bea
F^i_{\mu\nu} &=& \epsilon_{\mu\nu\lambda} \frac{2\pi i}{k_i e_i} {J}^{\lambda}.
\eea
and then we get
\bea
-i e_i F^i_{\mu\nu} &=& \epsilon_{\mu\nu\lambda} \sum_i \frac{2\pi}{k_i} {J}^{\lambda}
\eea
By identifying this with (\ref{close}) we find the constraint 
\bea
2\pi \sum_i \frac{1}{k_i} &=& -\lambda
\eea
on the abelian Chern-Simons levels $k_i$. 

For gauge group $U(1)\times U(1)$ we have $\lambda = 0$ and $f^{\bb\cc}{}_{\dd\aa}=0$ (vanishing structure constant is indeed implied by antisymmetry constaint when the three-algebra index $\aa$ takes only one value). Still we may describe this theory in the three-algebra formalism described above. We then find the constraint 
\bea
k_1 &=& -k_2
\eea
on the two Chern-Simons levels. The covariant derivative is given by 
\bea
D_{\mu} Z &=& \partial_{\mu} Z - i e_i A^i_{\mu} Z
\eea

In general $\lambda\neq 0$. The value of $\lambda$ is determined from the constraint 
\bea
f^{\bb\cc}{}_{\dd\aa} &=& - f^{\cc\bb}{}_{\dd\aa}.
\eea
when we decompose the structure constants as
\bea
f^{\bb\cc}{}_{\dd\aa} &=& \t f^{\bb\cc}{}_{\dd\aa} + \lambda \delta^{\bb}_{\aa} \delta^{\cc
}_{\dd}.
\eea

\subsection{Semi-simple times abelian factors}
We assume that the generators of semi-simple gauge group factors are traceless. We then find that
\bea
\t f^{\bb\cc}{}_{\dd\aa} &=& - \sum_{l=1}^L \frac{2\pi}{k_l} G^{A_lB_l} (t_{A_l})^{\bb}{}_{\aa
} (t_{B_l})^{\cc}{}_{\dd}
\eea
Here 
\bea
G_{A_l B_l} &=& (t_{A_l})^{\bb}{}_{\aa} (t_{B_l})^{\aa}{}_{\bb}
\eea
and $G^{A_l B_l}$ denotes its inverse. Here $k_l$ are related to the Chern-Simons level of simple group $G_l$ whose Lie algebra generators in the fundamental representation are $(t_{A_l})^{b_l}{}_{a_l}$ and are sitting in the generators of the whole semi-simple group $G=G_1\times \cdots \times G_L$ as
\bea
(t_{A_l})^{\bb}{}_{\aa} &=& \delta^{b_1}_{a_1} \cdots (t_{A_l})^{b_l}{}_{a_l} \cdots \delta^{b_L}_{a_L}.
\eea
We then see that
\bea
G_{A_l B_l} &=& \frac{M}{M_l} g_{A_l B_l}
\eea
where
\bea
M_l &=& \delta^{a_l}_{a_l},\cr
M &=& M_1 \cdots M_L,\cr
g_{A_l B_l} &=& (t_{A_l})^{b_l}{}_{a_l} (t_{B_l})^{a_l}{}_{b_l}.
\eea
The true Chern-Simons levels are now given by 
\bea
K_l &=& k_l \frac{M}{M_l}
\eea
and the structure constants become
\bea
\t f^{\bb\cc}{}_{\dd\aa} &=& - \sum_{l=1}^L \frac{2\pi}{K_l} g^{A_lB_l} (t_{A_l})^{\bb}{}_{\aa
} (t_{B_l})^{\cc}{}_{\dd}
\eea
where $g^{A_l B_l}$ denotes the inverse of $g_{A_l B_l}$.

The gauge fields $A^A$ and $A^{i}$ in the semi-simple and abelian gauge group factors respectively, are uniquely determined by $\bar{A}^b{}_a$ and associated Chern-Simons levels. For the semi-simple part we have
\bea
\t B^{\bb}{}_{\aa} &=& B^{\bb}{}_{\aa} \bar{f}^{\bb\cc}{}_{\dd\aa}
\eea
where
\bea
B^{\bb}{}_{\aa} &=& i\sum_l \frac{k_l}{2\pi} A^{A_l} (t_{A_l})^{\bb}{}_{\aa} 
\eea
The covariant derivative (\ref{derivative}) then becomes
\bea
D_{\mu} Z_{\aa} &=& \partial_{\mu}Z_{\aa} - i A^A (t_A)^{\bb}{}_{\aa} Z_{\bb} - i A^{i}_{\mu} e_{i} Z_{\aa}\cr
&=& \partial_{\mu}Z_{a_1\cdots a_L} - i \sum_l A^{A_l} (t_{A_l})^{b_l}{}_{a_l} Z_{a_1 \cdots b_l \cdots a_L} - i A^i_{\mu} e_i Z_{a_1 \cdots a_L}.
\eea
The Chern-Simons terms (\ref{cs}) plus (\ref{cs-ab}) becomes
\bea
&& \sum_l \frac{k_l}{4\pi} \tr_l\(AdA+\frac{2i}{3}A^3\) + \frac{k_i e_i^2}{4\pi} A^{i} dA^{i}\cr
&=& \sum_l \frac{K_l}{4\pi} \tr_l\(AdA+\frac{2i}{3}A^3\) + \frac{k_i e_i^2}{4\pi} A^{i} dA^{i}.
\eea
where in the first line the trace is over the gauge field components 
\bea
\delta^{b_1}_{a_1}\cdots A^{b_l}{}_{a_l}\cdots \delta^{b_L}_{a_L}
\eea
and in the second line the trace is over the components 
\bea
A^{b_l}{}_{a_l}.
\eea
The former trace is a factor $M/M_l$ times the latter trace, which is absorbed by the same factor in the relation $K_l = (M/M_l) k_l$. Defining gauge parameter 
\bea
\bar{\Lambda}^{\bb}{}_{\aa} &=& -i\Lambda^A (t_A)^{\bb}{}_{\aa} -i \Lambda^i e_i \delta^{\bb}_{\aa}
\eea
the gauge variation (\ref{gaug}) becomes
\bea
\delta Z_{\aa} &=& -i\Lambda^A (t_A)^{\bb}{}_{\aa} Z_{\bb} - ie^i\Lambda^i Z_{\aa},\cr
\delta A^A_{\mu} &=& -D_{\mu} \Lambda^A,\cr
\delta A^i_{\mu} &=& -D_{\mu} \Lambda^i.
\eea

\subsection{Discrete identification}
For gauge group $U(1)/\mb{Z}_N$ we identify
\bea
e^{\frac{2\pi i}{N}} \sim 1
\eea
Consequently the Dirac charge quantization condition becomes relaxed to 
\bea
e g &=& \frac{1}{N} \mb{Z}
\eea
where 
\bea
\int F &=& 2\pi g.
\eea
In this case the Chern-Simons action 
\bea
\frac{K Ne^2}{4\pi} \int_{S^1\times S^2} A dA
\eea
is well-defined modulo $2\pi$ for any $K\in \mb{Z}$. This is so because 
\bea
e\int_{S^2} F \in \frac{2\pi}{N} \mb{Z}
\eea
Then the action becomes
\bea
K e \int_{S^1} A
\eea
which is well-defined modulo $2\pi$ for any $K\in \mb{Z}$.

We now specialize to each of the gauge groups that were obtained in the general classification of $\N=6$ theories in \cite{Schnabl:2008wj} with particular attention to $U(1)$ factors and their Chern-Simons levels.

\subsection{Gauge group $SU(N)_K\times U(1)^r$}
Here we have structure constants
\bea
f^{bc}{}_{da} &=& -\frac{2\pi}{K}G^{AB}(t_A)^b{}_a (t_B)^c{}_d + \lambda \delta^b_a \delta^c_d.
\eea
The $SU(N)$ generators have the property
\ben
G^{AB} (t_A)^b{}_a (t_B)^c{}_d &=& \delta^b_d\delta^c_a-\frac{1}{N}\delta^b_a\delta^c_d.\label{su}
\een
With this we find 
\bea
f^{bc}{}_{da} &=& -\frac{2\pi}{K} \delta^b_d \delta^c_a + \(\frac{2\pi}{KN} + \lambda\) \delta^b_a \delta^c_d.
\eea
We find antisymmetric structure constants only for 
\bea
\lambda &=& \frac{2\pi}{K} \(1-\frac{1}{N}\).
\eea
We should now try and solve the constraint equation
\bea
2\pi \sum_i \frac{1}{k_i} &=& -\frac{2\pi}{K} \(1-\frac{1}{N}\)
\eea
on the Chern-Simons levels in the various $U(1)$ factors. For $N>2$ we find no integer valued Chern-Simons levels and so gauge invariance is broken in these theories with just one $U(1)$ factor. However for $N=2$ we can solve this constraint with just one $U(1)$ factor. In this case we find 
\bea
k_1 &=& -2K.
\eea
This corresponds to $U(2) = SU(2) \times U(1)/{\mb{Z}_2}$ gauge group. 

For $N>2$ we can solve the constraint equation by including two $U(1)$ factors. We then get
\bea
k_1 &=& -K,\cr
k_2 &=& KN.
\eea
This theory corresponds to gauge group $U(N)_K\times U(1)_{-K}$. Indeed the level $KN$ is what we should have in $U(1)/\mb{Z}_N$ gauge group that sits in $U(N)$ as 
\bea
U(N) = SU(N) \times U(1) / \mb{Z}_N
\eea
It seems like the constraint equation together with gauge invariance tells us how many $U(1)$'s we shall include along with any discrete identifications on the gauge group!

\subsection{Gauge group $Sp(N)_K\times U(1)^r$}
In the appendix we demonstrate that 
\ben
G^{AB} (t_A)^b{}_a (t_B)^c{}_d &=& \frac{1}{2} \({(J^{-1})}^{bc} J_{ad} + \delta^b_d \delta^c_a\)\label{sp-identity}
\een
for the generators in $Sp(N)$ where we use a convention where $Sp(1)=SU(2)$. Here $J_{ab}$ is the invariant tensor in $Sp(N)$ and indices range as $a=1,...,2N$. We can check that the identity is reasonable by restricting to $N=1$ where it gives the corresponding identity for $SU(2)$, and we prove it in general in the appendix \ref{sp1} by noting three relations are obtained by tracing over indices in three different ways (one of which uses the $J$ tensor) and by making a general three-parameter ansatz. 

The three-algebra structure constants are
\bea
f^{bc}{}_{da} &=& -\frac{2\pi}{K} G^{AB} (t_A)^b{}_a (t_B)^c{}_d + \lambda \delta^b_a \delta^c_d.
\eea
These become antisymmetric only for 
\bea
\lambda &=& \frac{\pi}{K}.
\eea
We can solve the constraint equation
\bea
2\pi \sum_i \frac{1}{k_i} &=& -\frac{\pi}{K}
\eea
by taking just one $U(1)$ factor and Chern-Simons level
\bea
k_1 &=& -2K
\eea
For $N=1$ this descends to the $U(2)$ theory we found above.

\subsection{Gauge group $SU(N)_K\times SU(M)_{-K}\times U(1)^r$}
We have the generators 
\bea
(t_A)^{b\b}{}_{a\a} &=& (t_A)^b{}_a \delta^{\b}_{\a},\cr
(t_{\hat{A}})^{b\b}{}_{a\a} &=& \delta^b_a (t_{\hat{A}})^{\b}{}_{\a}
\eea
in $SU(N)\times SU(M)$. We denote any $U(1)$ generator as
\bea
E^{b\b}{}_{a\a} &=& \delta^b_a \delta^{\b}_{\a}.
\eea
The three-algebra structure constants are (suppressing the heavy bifundamental index structure, for instance writing $f$ in place of $f^{b\hat{b}c\hat{c}}{}_{d\hat{d}a\hat{a}} \equiv f^{\bb\cc}{}_{\dd\aa}$)
\bea
f &=& -2\pi \(\frac{1}{K} g^{AB} t_A \otimes t_B - \frac{1}{K} g^{\hat{A}\hat{B}} t_{\hat{A}} \otimes t_{\hat{B}}\) + \lambda E \otimes E
\eea
We find that the structure constants become antisymmetric only for 
\bea
\lambda &=& -\frac{2\pi}{K}\(\frac{1}{N}-\frac{1}{M}\).
\eea
and the constraint equation becomes
\bea
2\pi \sum_i \frac{1}{k_i} &=& \frac{2\pi}{K}\(\frac{1}{N}-\frac{1}{M}\)
\eea
This suggests we take two $U(1)$ factors and the levels 
\bea
k_1 &=& KN,\cr
k_2 &=& -KM.
\eea
The quantization of these levels in turn tells us that we are really dealing with the gauge group $U(N)_K \times U(M)_{-K}$. 

We notice that the three-bracket of three scalar fields
\bea
[Z^A,Z^B;Z^C]_{a\a} &=& Z^A_{b\b} Z^B_{b\b} Z_C^{d\d} f^{b\b c\c}{}_{d\d a\a}
\eea
when inserting the explict solution for the structure constants,
\ben
f^{b\b c\c}{}_{d\d a\a} &=& -\frac{2\pi}{K} \(\delta^{b}_{d} \delta^c_a \delta^{\b}_{\a} \delta^{\c}_{\d} - \delta^b_a \delta^c_d \delta^{\b}_{\d} \delta^{\c}_{\a}\)\label{h}
\een
becomes
\bea
= \frac{2\pi}{K} \(Z^A_{a\b} Z_C^{\b c} Z^B_{c \a} - Z^B_{a\b} Z_C^{\b c} Z^A_{c \a}\)
\eea
which is indeed a matrix realization of the three-algebra. 

A curious fact is that in all cases the three-algebra structure constants were found to be real valued, despite nothing in principle prevent them from having components being complex numbers, such that the whole structure constant is subject to a hermiticity constraint.

\section{Monopoles for supersymmetry enhancement}
To enhance supersymmetry from $\N=6$ to $\N=8$ we need a monopole that transforms as a rank-2 tensor \cite{Gustavsson:2009pm}, 
\bea
W_{\aa\bb}
\eea
Moreover this monopole $W_{\aa\bb}$ shall be electrically charged under all the $U(1)$'s, with electric charges that are twice the charges of the scalar field $Z_{\aa}$. So it shall have electric charges 
\bea
2 e_i.
\eea
This monopole can now be used to bring some field $Z^{\aa}$ having electric charges $-e_i$ into a field $Z_{\aa} \equiv W_{\aa\bb} Z^{\bb}$ that has electric charges $-e_i+2e_i = +e_i$. 

We expressed the monopole in terms of three-algebra indices. These are normally bifundamental indices, $\aa=(a,\hat{a})$. In terms of these indices, the monopole is expressed as 
\bea
W_{ab}^{\hat{a}\hat{b}}.
\eea
 
The simples example of a monopole operator is for $SO(4)$ gauge group. Here we have
\bea
W_{\aa\bb} &=& \delta_{\aa\bb}.
\eea
where $\aa$ is a vector index of $SO(4)$. In terms of bifundamental indices of $SU(2)\times SU(2)$ this can be written as
\bea
W_{ab}^{\hat{a}\hat{b}} &=& \epsilon_{ab} \epsilon^{\hat{a}\hat{b}}.
\eea
We have no direct proof that this is actually a monopole operator. We notice that the delta function is covariantly constant,
\bea
D_{\mu} W_{\aa\bb} &=& 0.
\eea

We now ask whether monopole operators in such a representation can exist in certain ABJM theories for certain Chern-Simons levels. This existence is a necessary condition for supersymmetry enhancement to $\N=8$. It may not be a sufficient condition, and one has to assure that additional properties are obeyed. These are 
\begin{enumerate}
\item vanishing scaling dimension of the monopole
\item the monopoles shall close the $\N=8$ supersymmetry variations on-shell 
\end{enumerate}
For $U(N)\times U(N)$ gauge group, the existence of monopole operators with vanishing scaling dimension was shown in \cite{Benna:2009xd}.

\subsection{Gauge representation of monopole operator}
The main result \cite{Moore:1989yh} that we wish to now explain is that in Chern-Simons theory a GNO monopole \cite{Goddard:1976qe} with magnetic weights $\beta^I$, transforms in a representation of the gauge group corresponding to the highest weight 
\bea
\alpha_I &=& K G_{IJ} \beta^J
\eea
for any simple gauge group factor associated to level $K$. Here $G_{IJ}$ is the projection of $G_{AB}$ to the Cartan subalgebra. For the $U(1)_i$ factor (with $i$ labeling the various $U(1)$'s) we find the electric charge
\bea
\alpha_i &=& k_i n_i e_i
\eea
for a monopole with magnetic charge 
\bea
\int F &=& \frac{n_i}{e_i}
\eea
where $n_i \in \mb{Z}$. Here $k_i\in \mb{Z}$ denotes the Chern-Simons theory level.

We first study abelian gauge group. Let us make a gauge variation
\bea
\delta A &=& d \Lambda.
\eea
Let us consider Chern-Simons action $\L_{CS}=\frac{k e^2}{4\pi}\int_M A \wedge dA$ on a three-manifold $M$ with boundary $\partial M$. Its variation is given by
\bea
\delta \L_{CS} &=& \frac{ke^2}{4\pi} \int_{\partial M} \Lambda dA
\eea
Let us choose the boundary to be a cylinder along the $z$-axis with angular coordinate $\varphi \in [0,2\pi]$. Then this variation can be expressed as
\bea
\delta \L_{CS} &=& \frac{ke^2}{4\pi} \int \Lambda \(\partial_z A_{\varphi} - \partial_{\varphi} A_z\) dz \wedge d\varphi
\eea
To make this Chern-Simons action gauge invariant it has to be supplemented by a boundary term. We take this boundary term to be 
\bea
\L_{bndry} &=& \frac{k e^2}{4\pi} \int A_z A_{\varphi} dz \wedge d\varphi
\eea
Let us now consider a singular gauge variation, with gauge parameter
\bea
\Lambda &=& \frac{n}{e}\varphi
\eea
This makes the gauge parameter
\bea
e^{ie\Lambda} &=& e^{in\varphi}
\eea
single-valued as we encircle the $z$-axis for $n\in \mb{Z}$. But we no longer find that the sum $\L_{CS}+\L_{bndry}$ is invariant under this singular gauge variation. Instead of finding a cancelation between the two terms, we now find that $\delta L_{CS} = \delta L_{bndry}$ and they add up to
\bea
\delta \(\L_{CS} + \L_{bnrdy}\) &=& kne\int A_z dz
\eea
This means that a 't Hooft operator creating this singular gauge variation changes the exponentiated action $e^{iS}$ (including the necessary boundary term) by the amount of a Wilson line $W_{\alpha} = e^{i\alpha\int A}$ with electric charge
\bea
\alpha &=& kne.
\eea

All this has a counterpart for semi-simple gauge group. Here we have the Chern-Simons action $\L_{CS} = \frac{k}{4\pi} \int_M \tr\(AdA + \frac{2i}{3}A^3\)$ where the trace is taken over the fundamental representation. In the presence of a boundary cylinder this should be supplemented by a boundary term \cite{Elitzur:1989nr}
\bea
\L_{bndry} &=& \frac{k}{4\pi} \int_{\partial M} \tr\(A_z A_{\varphi}\)dz \wedge d\varphi
\eea
Making a singular gauge transformation with gauge parameter 
\bea
g &=& e^{i\beta^I H_I \varphi}
\eea
associated to magnetic weights $\beta^I$ (which are such that this gauge group element becomes single-valued as we encircle the $z$-axis) that are contracted with Cartan generators $H_I$, we again find equal contributions from $\delta \L_{CS}$ and $\delta L_{bndry}$ and the sum is
\bea
\delta \(\L_{CS} + \L_{bnrdy}\) &=& ik\int \tr\(\beta^I H_I A_z\) dz
\eea
We expand the gauge field as $A_z = A_z^I H_I + A_z^a E_a$ where $H_I$ are Cartan generators and $E_a$ are step operators, and we have the metric
\bea
G_{IJ} &=& \tr(H_I H_J)
\eea
and we can recast this variation in the form
\bea
\delta \(\L_{CS} + \L_{bnrdy}\) &=& ikG_{IJ}\beta^I \int A^J_z dz
\eea
This shows that a 't Hooft operator creating a monopole with magnetic weights $\beta^I$ corresponds to a Wilson line transforming in a representation with highest weight vector
\bea
\alpha_I &=& k G_{IJ}\beta^J
\eea
To see this, we note a formula for a closed Wilson loop \cite{Diakonov:1989fc}, 
\bea
\tr_R P \exp i \int A &=& \int Dg \exp i \alpha_I \int A^{(g),I}
\eea
Here $\alpha_I$ are the highest weight in the representation $R$ of the Wilson loop. To make the right-hand side gauge invariant we integrate over all non-singular gauge transformations parametrized by $g$. A corresponding formula exists for the open Wilson loop.

We now return to the issue of the existence of suitable monopoles for various ABJM gauge groups. We go through each gauge group in turn and start by $U(1)\times U(1)$ for pedagogical reasons.

\subsection{Gauge group $U(1)\times U(1)$}
We assume some non-vanishing electric charges $e_i$ for $i=1,2$. Conventionally they are fixed to be $+1$ and $-1$, and perhaps this can always be achieved by a suitable choice of units. We will however keep these charges arbitrary but non-vanishing. Then Dirac charge quantization, i.e. single-valuedness of the 't Hooft operator gauge parameters
\bea
g_i &=& e^{ie_i g_i \varphi} 
\eea
as we encircle the $z$-axis, amounts to magnetic charges
\bea
g_i &=& \frac{n_i}{e_i}
\eea
where $n_i\in \mb{Z}$. Corresponding electric charges are
\bea
\alpha_i &=& k_i n_i e_i.
\eea
Supersymmetry enhancement requires monopoles with charges 
\bea
\alpha_i &=& 2e_i
\eea
We should now also recall that the levels for this gauge group are constrained by
\bea
k_1 + k_2 &=& 0
\eea
(since $\lambda = 0$ in this case). The only integer solutions to these equations are 
\bea
n_i &=& k_i,\cr
k_1 &=& 2,\cr
k_2 &=& -2
\eea
and 
\bea
n_i &=& 2k_i,\cr
k_1 &=& 1,\cr
k_2 &=& -1.
\eea
This is to say that we can expect $\N=8$ supersymmetry only for $k_1=1,2$. Indeed this is true and can be verified explicitly. We may for instance express the theory as a sigma model on $\mb{C}^4/\mb{Z}_{k_1}$ and then it is clear that this orbifold preserves $SO(8)$ for $k_1=1,2$ only.

\subsection{Gauge group $U(N)\times U(M)$}
We can limit our study to just the $U(N)$ factor of the gauge group since the $U(M)$ factor will tell a similar story. There is no need to assume that $N=M$. Let us choose Cartan generators in $U(N)$ as
\bea
T_I &=& \diag(0,...,1,....,0)
\eea
We can then easily solve the Dirac quantization condition
\bea
e^{2\pi i\gamma^I T_I} &=& 1
\eea
for the magnetic weights $\gamma^I$, and we find a smallest monopole charge $\gamma^I T_I$ being of the form 
\bea
\(\begin{array}{cccc}
1 & 0 & ... & 0\\
0 & 0 & ... & 0\\
  &   & \ddots & \\
0 & 0 & ... & 0
\end{array}\) &=& \frac{1}{N}\(\begin{array}{cccc}
N-1 & 0 & ... & 0\\
0 & -1 & ... & 0\\
  &   & \ddots & \\
0 & 0 & ... & -1
\end{array}\)\cr
&& + 
\frac{1}{N}\(\begin{array}{cccc}
1 & 0 & ... & 0\\
0 & 1 & ... & 0\\
  &   & \ddots & \\
0 & 0 & ... & 1
\end{array}\)
\eea
Here we have separated the monopole charge into a traceless and a trace part, corresponding to a separation
\bea
\gamma^I T_I &=& \beta^I H_I + \beta e
\eea
where $H_I$ are Cartan generators in $SU(N)$, and $e$ an electric charge of $U(1)$. It is convenient to take the $H_I$ to be orthogonal, and normalize them as
\bea
\tr (H_I H_J) &=& \frac{1}{2} \delta_{IJ}
\eea
in the fundamental representation $N$ of $SU(N)$. We may choose them as
\bea
H_I &=& \frac{1}{\sqrt{2I(I+1)}} \diag(\underbrace{1,1,...,1}_{I},-I,0,...,0)
\eea
for $I=1,...,N-1$. We can then read off the magnetic weight as
\bea
\beta^I &=& 2\(0,0,...,0,-\frac{1}{\sqrt{2}}\sqrt{1-\frac{1}{N}}\),\cr
\beta &=& \frac{1}{Ne}.
\eea
With our choice of $H_I$ the weights in the fundamental representation of $SU(N)$ are given by 
\bea
w_1 &=& \frac{1}{\sqrt{2}} \(\frac{1}{\sqrt{1.2}},\frac{1}{\sqrt{2.3}},...,\frac{1}{\sqrt{(N-1)N}}\),\cr
w_2 &=& \frac{1}{\sqrt{2}} \(-\frac{1}{\sqrt{1.2}},\frac{1}{\sqrt{2.3}},..., \frac{1}{\sqrt{(N-1)N}}\),\cr
&&...\cr
w_N &=& \frac{1}{\sqrt{2}} \(0,0,....,-\sqrt{1-\frac{1}{N}}\)
\eea
According to our general result, the highest weight of the representation of the monopole is given by 
\bea
\alpha_I &=& K G_{IJ} \beta^J.
\eea
For the abelian part we have the electric charge
\bea
\alpha &=& k \beta e^2.
\eea
which follows from our general result $\alpha = kne$ by taking magnetic charge $\beta=n/e$. We have the abelian level 
\bea
k &=& KN
\eea
Recalling that $\beta=1/(Ne)$ for our smallest monopole, we get
\bea
\alpha &=& K e
\eea
To enhance supersymmetry we want the charge
\bea
\alpha &=& 2e
\eea
so we shall take the level to be 
\bea
K=2
\eea
with this monopole. We can also double the monopole charge and take $K=1$. We can never get enhancement of supersymmetry for $K>2$. Now we must also consider the the $SU(N)$ representation. We take $K=2$ and get a cancelation of factor $2$ from $K$ and another factor $\frac{1}{2}$ from $G_{IJ}$, so that
\bea
\alpha_I &=& 2 (w_N)_I
\eea
which corresponds to a symmetric rank-2 tensor.

It is interesting to note that neither the $SU(N)$ part
\bea
e^{2\pi i \beta^i H_i} 
\eea
nor the $U(1)$ part
\bea
e^{2\pi i \beta \frac{1}{Ne}}
\eea
of the gauge group element, equals unity, and only their product obeys Dirac charge quantization condition
\bea
e^{2\pi i \beta^i H_i} e^{2\pi i \beta \frac{1}{Ne}} &=& 1.
\eea

\subsection{Gauge group $SU(N)\times SU(N)$}
With no $U(1)$'s we must take equal ranks in the two gauge group factors, that is $SU(N)\times SU(M)$ with $N=M$. For gauge group $SU(N)\times SU(N)$ the smallest magnetic charge that obeys Dirac charge quantization is of the form
\bea
\diag(1,1,...,1,-N+1) &=& \sqrt{2 N(N-1)} H_{N-1}
\eea
for the left $SU(N)$ factor, and similarly for the right $SU(N)$. Now this corresponds to magnetic weight
\bea
\beta^I &=& \(0,0,...,0,\sqrt{2 N (N-1)}\)\cr
&=& -2N (w_N)^I.
\eea
We then get the highest weight for the corresponding gauge representation as
\bea
\alpha_I &=& K G_{IJ} \beta^J\cr
&=& KN (w_N)_I
\eea
which is a factor $N$ too large. Another way to see the factor of $N$ is by recalling that $\beta$ are weights in the Langlands dual of the gauge group \cite{Goddard:1976qe}, which in this case is given by $SU(N)/{\mb{Z}_N} \times SU(N)/{\mb{Z}_N}$. The weight lattice in this dual gauge group involves a scaling by $N$ of the corresponding weight lattice for $SU(N)\times SU(N)$, thereby explaining the factor of $N$. 

We conclude we can not find any monopole transforming like a symmetric rank-2 tensor for any $N>3$. Hence no supersymmetry enhancement to $\N=8$.

For $N=2$ we find a symmetric rank-2 representation only for $K=1$. This monopole gives a second copy of $\N=8$ supersymmetry charges, which differ from those in BLG theory. Having two copies of superconformal algebra may sound strange, but is justified if the theory is dual to $\N=8$ SYM with gauge group $U(2)$ \cite{Bashkirov:2010kz}. Indeed it was shown in \cite{Lambert:2010ji} that $U(2)\times U(2)$ ABJM theory at $K=1$ is dual to $(SU(2)\times SU(2))/{\mb{Z}_k}$ ABJM theory at $K=1$, and in \cite{Kapustin:2010xq}, \cite{Bashkirov:2010kz} much evidence was provided that implies that $U(2)\times U(2)$ ABJM theory at $K=1$ is dual to $U(2)$ $\N=8$ SYM.

\subsection{Gauge group $Sp(N)_K\times U(1)_{-2K}$}
Here we expect enhanced $\N=8$ supersymmetry for $K=1$ only. The moduli space is $\mb{C}^4/{\mb{Z}_{2K}}$ \cite{Aharony:2008gk}. 

We now show that we can find a monopole that transforms as a symmetric rank-2 tensor only for level $K=1$. We impose Dirac charge quantization conditions
\bea
e^{2\pi i \beta^i w_i} &=& 1,\cr
e^{2\pi i \beta e} &=& 1
\eea
Here 
\bea
w_i &=& \pm \frac{1}{2} e_i
\eea
are weights in $Sp(N)$ where $e_i$ is a vector with $N$ components which has entries being zero except for the entry at position $i$ that is one. This normalization of the weights is the one we find if we choose the following normalization for the metric 
\bea
G_{ij} &=& \frac{1}{2} \delta_{ij}.
\eea
For the $U(1)$ we assume an electric charge $e$. 

These charge quantization conditions imply
\bea
\beta^i e_i &=& 2m,\cr
\beta e &=& n
\eea
for some integer numbers $n$ and $m$. Demanding symmetric rank-2 representation means that we have highest weights
\bea
\alpha_i &=& 2 \frac{1}{2} e_i,\cr
\alpha &=& 2 e
\eea
We then get the equations
\bea
e_i &=& K G_{ij} \beta^j,\cr
2e &=& -2K e^2\beta
\eea
and they yield
\bea
Km &=& 1,\cr
Kn &=& -1
\eea
which we can solve with integer numbers $n$ and $m$ only for $K=1$.

Since level for $U(1)$ part is $-2K$ we can also consider gauge group $Sp(N)\times U(1)/{\mb{Z}_2}$. This amounts to Dirac quantization exp$2\pi i \beta e=\pm 1$ which results in equations
\bea
Km &=& 1,\cr
K\frac{n}{2} &=& -1
\eea
Again we only have integer solutions $m,n$ for $K=1$.

\section{$\N=5$ supersymmetric Chern-Simons theories}

The $\N=6,8$ theories can be thought of as obtained from $\N=5$ theories by restriction of the gauge group. The algebraic structure of $\N=5$ theories therefore unifies the $\N=5,6,8$ theories. In \cite{deMedeiros:2008zh} it was suggested that this algebraic structure should be the quaternionic three-algebras. The supersymmetry variations of certain $\N=5$ theories was obtained in \cite{Hosomichi:2008jb} and these have also been shown to have a quaternionic three-algebra description in \cite{Chen:2009cwa}, \cite{Bagger:2010zq}, \cite{Kim:2010kq}.

We may extend the analysis of supersymmetry enhancements to ABJM theories with $\N=5$ supersymmetry, which for certain low levels may have enhanced $\N=6$ supersymmetry. We will show that all $\N=5$ ABJM theories can be described by either one of two different languages: by hermitian three-algebra or by quaternionic three-algebra. Moreover there is a way to translate from one language to the other language. The hermitian three-algebra structure constants of $\N=5$ theories are constrained by exactly the same conditions as the structure constants of $\N=6$ ABJM theories. Recalling the relationship between three-algebra and Lie algebras, we conclude that all $\N=5$ theories are classified by the essentially the same gauge groups as the $\N=6$ ABJM theories (at least they share the same Lie algebras). If on the other hand we choose to describe $\N=5$ ABJM theories by quaternionic three-algebra, then we find different structure constants, and in turn different gauge groups \cite{Bergshoeff:2008bh}, \cite{Kim:2010kq}, \cite{Bagger:2010zq}. The gauge group is not a physical observable of the theory, and we think it is possible that different gauge groups can describe the same physics. 

\subsection{Quaternionic three-algebra}
We assume there is a gauge invariant symplectic form $\omega_{\qa\qb}$,
\bea
\omega_{\qa\qb}\omega^{\qb\qc} &=& -\delta_{\qa}^{\qb},\cr
\omega^{\qa\qb}\omega_{\qb\qc} &=& -\delta^{\qa}_{\qb},\cr
\omega_{\qa\qb} &=& -\omega_{\qb\qa},\cr
\bar{\Lambda}^{\qc}{}_{\qa} \omega_{\qc\qb} + \bar{\Lambda}^{\qc}{}_{\qa} \omega_{\qa\qc} &=& 0.
\eea
Here $\bar{\Lambda}^{\qa}{}_{\qb}$ denotes a gauge parameter where we use bar instead of tilde since tilde will be used for something different later on. We denote the inverse of the symplectic form as $(\omega^{-1})^{\qa\qb}$ and this is thus given by
\bea
(\omega^{-1})^{\qa\qb} &=& -\omega^{\qa\qb}.
\eea
We lower upper indices by contracting by $\omega_{\qa\qb}$ from the left, and we rise indices by contracting by $(\omega^{-1})^{\qa\qb}$ from the left. So we have
\bea
v_{\qa} &=& \omega_{\qa\qb} v^{\qb},\cr
v^{\qa} &=& (\omega^{-1})^{\qa\qb} v_{\qb}.
\eea
Gauge invariance of $\omega_{\qa\qb}$ implies a symmetric gauge parameter 
\bea
\bar{\Lambda}^{\qa\qb} &=& \bar{\Lambda}^{\qb\qa}.
\eea
We assume that we may express the gauge parameter as
\bea
\bar{\Lambda}^{\qb\qa} &=& \Lambda_{\qc\qd} g^{\qb\qc\qd\qa}
\eea
where $g^{\qb\qc\qd}{}_{\qa}$ are gauge invariant structure constants. Gauge invariance of these structure constants implies that they satisfy the fundamental identity. If we introduce generators $T^{\qa}$ and define structure constants as
\bea
\{T^{\qa},T^{\qb},T^{\qc}\} &=& g^{\qa\qb\qc}{}_{\qd} T^{\qd}
\eea
then the quaternionic fundamental identity that is satisfied by this three-bracket, reads
\bea
\{\{x,y,z\},u,v\} &=& \{\{x,u,v\},y,z\} + \{x,\{y,u,v\},z\} + \{x,y,\{z,u,v\}\}.
\eea
As we show in the appendix, the fundamental identity implies that 
\bea
g^{\qb\qc\qd\qa} &=& \kappa_{AB} (t^A)^{\qc\qd} (t^B)^{\qa\qb}.
\eea
where $\kappa_{AB}$ is a Killing form on the associated Lie algebra. It follows that the structure constants are subject to the symmetry
\bea
g^{\qb\qc\qd\qa} &=& g^{\qd\qa\qb\qc}.
\eea
From this it follows that $\bar{\Lambda}^{\qb\qa}$ is symmetric if and only if $\Lambda_{\qc\qd}$ is symmetric. This in turn implies the additional symmetry 
\bea
g^{\qb\qc\qd\qa} &=& g^{\qb\qd\qc\qa}.
\eea
A nice reference for quaternionic three-algebra is \cite{deMedeiros:2008zh}. Here it is shown that q-algebra and h-algebra (hermitian three-algebra) are equivalent. For any two elements $x$ and $y$ in the h-algebra the inner product is a complex number $h(x,y)$ being complex anti-linear in say its second entry. The q-algebra has the inner product
\bea
\omega(x,y) &=& h(x,\omega y).
\eea
where $\omega$ is a linear map such that
\bea
\omega^2 &=& -1,\cr
h(\omega x,\omega y) &=& h(y,x).
\eea
We use the same letter $\omega$ also for the symplectic product, the distinction between the two should be always clear from the context. We find that 
\bea
\omega(y,x) &=& -\omega(x,y).
\eea
If $T^{\qa}$ denote generators of a q-algebra we define
\bea
h(T^{\qa},T^{\qb}) &=& \delta^{\qa}_{\qb},\cr
\omega(T^{\qa},T^{\qb}) &=& \omega^{\qa\qb}
\eea
We define 
\bea
\omega_{\qb\qa}^* = (\omega^{-1})^{\qa\qb}.
\eea

The hermitian three-bracket (h-bracket) $[x,y;z]$ is complex linear in its first two entries and complex anti-linear in its third. We define the quaternionic three-bracket (q-bracket) as
\ben
\{x,y,z\} &=& [x,y;\omega z].\label{transition}
\een
From gauge invariance we have that the q-bracket is symmetric under exchange of $y$ and $z$. We do not assume the h-bracket is antisymmetric in $x$ and $y$. In the appendix we demonstrate that this antisymmetry is not needed in order to satisfy the hermitian fundamental identity
\bea
[[x,y;z],u;v] &=& [[x,u;v],y;z] + [x,[y,u;v];z] - [x,y;[z,v;u]].
\eea
It is easy to see that we can go from hermitian fundamental identity to quaternionic fundamental identity using (\ref{transition}). Both the q-bracket and h-brackets are subject to 
\bea
\{\omega X,\omega Y,\omega Z\} &=& \omega \{X,Y,Z\},\cr
[\omega X,\omega Y;\omega Z] &=& \omega [X,Y;Z]
\eea
For example using the matrix realization of the h-bracket we have
\bea
[\omega X,\omega Y;\omega Z] = \omega X (\omega Y)^{\dag} \omega Z - .. = \omega X Y^{\dag} Z - .. = \omega[X,Y;Z].
\eea 
The h-bracket satisfies the trace invariance condition (gauge invariance)
\bea
h([a,x;y],b) - h(a,[b,y;x]) &=& 0
\eea
For the q-bracket this amounts to
\bea
\omega(\{a,x,y\},b) + \omega(a,\{b,x,y\}) &=& 0.
\eea
Now since the q-bracket is symmetric one may suspect that we can also make that symmetry explicit, and thus define
\ben
\{x,y,z\} &=& [x,y;\omega z] + [x,z;\omega y]\label{qq}
\een
This suspicion turns out to be correct. If we assume that the h-bracket satisfies the hermitian fundamental identity, we can show that the q-bracket defined this way indeed satisfies the quaternionic fundamental identity. We demonstrate this explicitly in the Appendix \ref{qbracket}. Henceforth we shall always use this latter definition (\ref{qq}) of the q-bracket. 

If we have the relation (\ref{transition}), then (\ref{qq}) must follow.\footnote{This is up to a factor of $2$ which we may absorb in a rescaling the h-bracket in (\ref{qq}) by a factor of $2$ in comparison to the h-bracket in (\ref{transition}).} This is so, because the left-hand side of (\ref{transition}) is symmetric in $y,z$. So all we do in (\ref{qq}) is just a trivial rewriting of (\ref{transition}). Now we may invert (\ref{transition}) and get a h-bracket from the q-bracket as $[x,y;z] = -\{x,y,\omega z\}$. This will satisfy the hermitian fundamental identity if the q-bracket satisfies the quaternionic fundamental identity. However, the h-bracket in (\ref{qq}) can be more general than the h-bracket in (\ref{transition}) since we make a symmetry explicit which means that we can relax that symmetry of the bracket itself. It means that the h-bracket no longer really has to obey the symmetry that puts the two terms equal, i.e. $[x,y,\omega z] = [x,z;\omega y]$, may no longer hold in (\ref{qq}) even though it holds in (\ref{transition}). It may happen therefore happen that the h-bracket in (\ref{qq}) does not always satisfy the hermitian fundamental identity despite the q-bracket on the left-hand side satisfies the quaternionic fundamental identity.

\subsubsection{Matrix realization}
The q-algebra generators may be realized by matrices $(T^{\qa})_a{}^{\hat{a}}$. Their conjugates are denoted $(T_{\qa})_{\hat{a}}{}^a$. If the gauge group is $Sp(N)\times O(M)$ then $a = 1,...,2N$ and $\hat{a} = 1,...M$ and $T^{\qa}$ are $2N\times M$ matrices, while their conjugates $T_{\qa}$ are $M\times 2N$ matrices. We should choose these matrices to be normalized so that 
\bea
h(T^{\qa},T^{\qb}) = \tr(T^{\qa} T_{\qb}) = \delta^{\qa}_{\qb}.
\eea
We define matrices $(\t T_{\qa})_{a}{}^{\a} = \omega_{\qa\qb} (T^{\qb})_a{}^{\a}$ and $(\t T^{\qa})_{\a}{}^a = (\omega^{-1})^{\qa\qb} (T_{\qb})_{\a}{}^a$. 

The matrix realization of the h-bracket is
\bea
[T^{\qa},T^{\qb};T^{\qc}] &=& \alpha T^{\qa} T_{\qc} T^{\qb} + \beta T^{\qb} T_{\qc} T^{\qa}\label{mat}
\eea
This satisfies the hermitian fundamental identity for any choice of parameters $\alpha$ and $\beta$ (see the Appendix). With this h-bracket, we find the matrix realization of the q-bracket as
\bea
\{T^{\qa},T^{\qb},T^{\qc}\} &=& \alpha \(T^{\qa} \t T^{\qc} T^{\qb} + T^{\qa} \t T^{\qb} T^{\qc}\) + \beta\(T^{\qb} \t T^{\qc} T^{\qa} +  T^{\qc} \t T^{\qb} T^{\qa}\).\label{matrix}
\eea

\subsubsection{Gauge groups associated with three-algebras}

The gauge group has to be generated by symmetric Lie algebra generators. One possibility is to take gauge group $Sp(N)_K\times SO(M)_{2L}$ and associated levels $K$ and $L$. We note that the Chern-Simons level in $SO(2)$ gauge group is twice the level of $U(1)$ gauge group, coming from replacing the imaginary unit by a $2\times 2$ matric that squares to minus one. For the fundamental (vector) representation of $SO(M)$ we find
\bea
G^{\hat{A}\hat{B}} (t_{\hat{A}})^{\hat{b}}{}_{\hat{a}} (t_{\hat{B}})^{\hat{c}}{}_{\hat{d}} &=& \frac{1}{2}\(\delta^{\hat{b}\hat{c}} \delta_{\hat{a}\hat{d}} - \delta^{\hat{c}}_{\hat{a}} \delta^{\hat{b}}_{\hat{d}}\)
\eea
Using this and the corresponding result for $Sp(N)$, we find the q-algebra structure constants
\bea
f^{\bb\hat{b}\cc\hat{c}}{}_{\dd\hat{d}\aa\hat{a}} &=& -\frac{\pi}{K}\({(J^{-1})}^{\bb\cc}J_{\aa\dd}+\delta^{\bb}_{\dd}\delta^{\cc}_{\aa}\)\delta^{\hat{b}}_{\hat{a}} \delta^{\hat{c}}_{\hat{d}} - \frac{\pi}{L} \delta^{\bb}_{\aa} \delta^{\cc}_{\dd} \(\delta^{\hat{b}\hat{c}}\delta_{\hat{a}\hat{d}} - \delta^{\hat{c}}_{\hat{a}} \delta^{\hat{b}}_{\hat{d}}\)
\eea
Rising indices by 
\bea
(\omega^{-1})^{\qa\qb} &=& {(J^{-1})}^{\aa\bb} \delta^{\hat{a}\hat{b}}
\eea
we get
\bea
f^{\bb\hat{b}\cc\hat{c}\dd\hat{d}\aa\hat{a}} &=& \frac{\pi}{K}\({(J^{-1})}^{\bb\cc}{(J^{-1})}^{\aa\dd}-{(J^{-1})}^{\aa\cc}{(J^{-1})}^{\dd\bb}\)\delta^{\hat{b}\hat{a}}\delta^{\hat{c}\hat{d}} \cr
&&-\frac{\pi}{L} {(J^{-1})}^{\aa\bb}{(J^{-1})}^{\dd\cc} \(\delta^{\hat{b}\hat{c}} \delta^{\hat{a}\hat{d}} - \delta^{\hat{c}\hat{a}} \delta^{\hat{b}\hat{d}} \)
\eea
We note that this has the desired symmetry
\bea
f^{\qb \qc \qd \qa} &=& f^{\qb \qd \qc \qa}.
\eea

If we define 
\bea
(\t Z^A)^{\a \aa} &=& {(J^{-1})}^{\aa\bb} \delta^{\a\b} Z^A_{\bb\b} 
\eea
we find that the matrix realization of the three-bracket with these structure constants become
\bea
\{X,Y,Z\} &=& -\frac{\pi}{K} \(X\t Y Z + X \t Z Y\) + \frac{\pi}{L} \(Z \t Y X + Y \t Z X\).
\eea
We then also identify the h-bracket as
\bea
[X,Y;Z] &=& -\frac{\pi}{K} X Z^{\dag} Y + \frac{\pi}{L} Y Z^{\dag} X
\eea
This is antisymmetric only of $K=L$ and this is the case that was studied in \cite{Hosomichi:2008jb}. It seems like $\N=5$ supersymmetric theories can only exist when $K=L$ even though this constraint does not follow direcly from symmetry constraint $f^{\qb\qc\qd\qa} = f^{\qb\qd\qc\qa}$, but it does follow from a slightly more intricate symmetry constraint, namely the condition that there exists a decomposition according to Eq (\ref{qq})
\ben
f^{\qb\qc\qd\qa} &=& g^{\qb\qc\qd\qa} + g^{\qb\qd\qc\qa}\label{split}
\een
such that 
\ben
g^{\qb\qc\qd\qa} &=& - g^{\qc\qb\qd\qa}.\label{as}
\een
The antisymmetry (\ref{as}) is an additional constraint we put on the q-algebra.\footnote{We can make the symmetry explicit,
\bea
f^{\qb\qc\qd\qa} &=& \frac{1}{2} \(f^{\qb\qc\qd\qa} + f^{\qb\qd\qc\qa}\)
\eea
and then we can also make the antisymmetry manifest, at the cost of adding a term as
\bea
&=& \frac{1}{4}\((f^{\qb\qc\qd\qa}-f^{\qc\qb\qd\qa}) + (f^{\qb\qd\qc\qa} - f^{\qd\qb\qc\qa})\) + \frac{1}{4}\(f^{\qc\qb\qd\qa}+f^{\qd\qb\qc\qa}\)
\eea
We see that the decomposition (\ref{split}) with the requirement (\ref{as}) is unique and exists only if the structure constants are subject to the antisymmetry
\bea
f^{\qc\qb\qd\qa}+f^{\qd\qb\qc\qa} &=& 0.
\eea} It is not needed to define a q-algebra, but it seems to be needed in order to have $\N=5$ supersymmetry. This is analogous to the antisymmetric constraint $f^{\bb\cc}{}_{\dd\aa} = - f^{\cc\bb}{}_{\dd\aa}$ that Bagger and Lambert put on the h-algebra structure constants. This is not needed to define a h-algebra\footnote{In Appendix \ref{halge} we show that the hermitian fundamental identity does not rely on that antisymmetry property.}, but it is needed for $\N=6$ supersymmetry.

For $K=L$ we can split the structure constants of $Sp(N)\times SO(M)$ gauge group according to Eq (\ref{split}) in such a way that we obtain
\bea
g^{\qb\qc\qd\qa} &=& -\frac{\pi}{K}\({(J^{-1})}^{\aa\bb} {(J^{-1})}^{\cc\dd} \delta^{\b\d} \delta^{\a\c} - {(J^{-1})}^{\aa\cc} {(J^{-1})}^{\bb\dd} \delta^{\a\b} \delta^{\c\d}\)
\eea
antisymmetric in $\qb\qc$. Lowering indices we get
\bea
2g^{\qb\qc}{}_{\qd\qa} &=& -\frac{2\pi}{K} \(\delta^{\bb}_{\aa} \delta^{\cc}_{\dd} \delta^{\b}_{\d} \delta^{\c}_{\a} - \delta^{\cc}_{\aa} \delta^{\bb}_{\dd} \delta^{\b}_{\a} \delta^{\c}_{\d}\)
\eea
We have now obtained the h-structure constants of $U(2N)_{K}\times U(M)_{-K}$ ABJM theory.\footnote{A factor of $2$ comes from the sum of two terms in Eq (\ref{split}) so the actual structure constant will be $2g^{\qb\qc}{}_{\qd\qa}$ rather than $g^{\qb\qc}{}_{\qd\qa}$.} Due to reality conditions on the fields we have $\N=5$ supersymmetry only, but for $K=1$ we expect $\N=6$ supersymmetry. The resulting $\N=6$ theory can not be $U(2N)\times U(M)$ $\N=6$ ABJM theory at level $K=1$ and with no reality conditions on matter fields. It should be some $\N=6$ ABJM theory with no reality conditions for some gauge group and some Chern-Simons level. This is expected if there are no other candiditate $\N=6$ theories to choose among. 

We do not have a unique description -- in particular no unique gauge group -- of $\N=5$ ABJM theory. The gauge group depends on the way we formulate the theory. If we formulate the theory in terms of q-algebra we may have a gauge group that is $Sp(N)\times SO(M)$, and if we formulate the same theory in terms of h-algebra the gauge group is $U(2N)\times U(M)$. The theory then has a gauge group of $\N=6$ ABJM theory, and we see that this holds in general. So the complete classification of $\N=5$ theories correspond precisely to the classification of $\N=6$ ABJM theories. The difference is that for $\N=5$ theories the matter fields are subject to reality conditions. These reality conditions in particular generally break $\N=6$ down to $\N=5$ supersymmetry. The gauge groups allowed are the same as those allowed for $\N=6$ if the theory is formulated in terms of h-algebra. 

Let us next consider the covariant derivative of $Sp(N)\times SO(M)$ gauge group,
\bea
D_{\mu} Z_{\qa} &=& \partial_{\mu} Z_{\qa} + A_{\mu,\qc\qd} f^{\qb\qc\qd}{}_{\qa} Z_{\qb}.
\eea
Since the gauge field is symmetric, we see that this can be written as
\bea
D_{\mu} Z_{\qa} &=& \partial_{\mu} Z_{\qa} + A_{\mu,\qc\qd} (2g)^{\qb\qc\qd}{}_{\qa} Z_{\qb}
\eea
and then we may rise and lower indices and get
\bea
D_{\mu} Z_{\qa} &=& \partial_{\mu} Z_{\qa} + A_{\mu}{}^{\qd}{}_{\qc} (2g)^{\qb\qc}{}_{\qd\qa} Z_{\qb}
\eea
which is the covariant derivate associated with gauge group $U(2N)\times U(M)$.

Given an h-algebra, there is an associated q-algebra. The simplest example is the h-algebra of $Sp(N)\times U(1)$ gauge group. This comes with h-structure constants
\bea
g^{\bb\cc}{}_{\dd\aa} &=& -\frac{\pi}{K}\({(J^{-1})}^{\bb\cc} J_{\aa\dd} + \delta^{\bb}_{\dd} \delta^{\cc}_{\aa} - \delta^{\bb}_{\aa} \delta^{\cc}_{\dd}\)
\eea
Rising indices with ${(J^{-1})}^{\aa\bb}$, which we may identify with the symplectic form on the q-algebra, we get the q-algebra structure constants
\bea
f^{\bb\cc\dd\aa} &=& -\frac{2\pi}{K}\({(J^{-1})}^{\bb\cc} {(J^{-1})}^{\aa\dd} + {(J^{-1})}^{\dd\bb} {(J^{-1})}^{\aa\cc}\).
\eea
These are nothing but q-algebra structure constant associated with gauge group $Sp(N)$. See Eq (\ref{sp-identity}) and rise indices by the symplectic form ${(J^{-1})}^{\aa\bb}$. 

We will give more details on the q-algebra and h-algebra formulations of $\N=5$ theories below.

\subsubsection{A comment on the Chern-Simons level}
We have on the one hand an $\N=6$ supersymmetric theory with gauge group $Sp(N) \times U(1)$. On the other hand we have an $\N=5$ supersymmetric theory with gauge group $Sp(N) \times SO(2)$. The amount of global symmetries being different, these theories can impossibly be dual to each other despite they have isomorphic gauge groups. 

So far we have used an overall definition of Chern-Simons levels -- we have defined the level as the quantity $K$ in the Chern-Simons term 
\bea
\frac{K}{4\pi} \tr\(A dA - \frac{2i}{3} A^3\)
\eea
where the trace is taken in the fundamental representation of the gauge group. Let us now specialize to $SO(2)$ gauge group. Then we have the gauge field
\bea
I A_{\mu} 
\eea
where
\bea
I &=& \(\begin{array}{cc}
0 & -1\\
1 & 0
\end{array}\)
\eea
acting on a matter field 
\bea
Z = X + I Y = \(\begin{array}{cc}
X & -Y\\
Y & X
\end{array}\)
\eea
and its conjugate
\bea
Z^T = X - IY
\eea
such that we have a covariant derivative
\bea
D_{\mu} Z &=& \partial_{\mu} Z - I A_{\mu} Z.
\eea
Corresponding $U(1)$ fields are $i A_{\mu}$ and $Z = X+iY$. We have the $SO(2)$ action
\bea
\frac{K}{4\pi} \tr (AdA) - \tr(D_{\mu} Z D^{\mu} Z^T)
\eea
and this equals the corresponding $U(1)$ action up to an overall factor of $2$
\bea
= 2 \(\frac{K}{4\pi} AdA - |D_{\mu} Z|^2\).
\eea
We can then make a field redefinition and we deduce that the Chern-Simons level $\hat{K}$ that we defined as above, actually is related to the $U(1)$ level as
\bea
K &=& 2\hat{K} 
\eea
and it is $K$ and not $\hat{K}$ that is integer quantized.

This generalizes to $SO(M)$ gauge groups for any $M$. The integer quantized Chern-Simons levels on these Chern-Simons gauge groups are thus $Sp(N)_K\times SO(M)_{-2K}$.

\subsection{Reality conditions on matter fields}\label{realcond}
To get $\N=5$ supersymmetry we must impose the reality conditions
\bea
(X_{A\qa})^* &=& (\Omega^{-1})^{AB}(\omega^{-1})^{\qa\qb}X_{B\qb}.
\eea
one all matter fields. Our convention is such that 
\bea
(\Omega_{AB})^* &=& -(\Omega^{-1})^{AB},\cr
(\omega_{\qa\qb})^* &=& -(\omega^{-1})^{\qa\qb}.
\eea
We can rewrite the above reality condition in the alternative form
\bea
(X^A_{\qa})^* &=& -(\omega^{-1})^{\qa\qb}\Omega_{AB}X^B_{\qb}.
\eea
We define a q-tensor as a quantity whose upper indices are rised by $\Omega^{-1}$ from the left to the right according to
\bea
T^{AB...} &=& (\Omega^{-1})^{AC} (\Omega^{-1})^{BD} ... T_{CD...}.\label{tensor}
\eea
The following argument is taken from \cite{Bagger}. Any antisymmetric q-tensor $T_{AB}$ with two indices can be decomposed into a traceless and a trace part as
\bea
T_{AB} &=& \hat{T}_{AB} + \Omega_{AB} T
\eea
with 
\bea
\Omega^{AB} \hat{T}_{AB} &=& 0.
\eea
If we define
\bea
\epsilon^{ABCD} &=& (\Omega^{-1})^{AB} (\Omega^{-1})^{CD} - (\Omega^{-1})^{AC} (\Omega^{-1})^{BD} - (\Omega^{-1})^{AD} (\Omega^{-1})^{CB}
\eea
then we find that 
\bea
\frac{1}{2} \epsilon^{ABCD} \hat{T}_{CD} &=& - \hat{T}^{AB},\cr
\frac{1}{2} \epsilon^{ABCD} \Omega_{CD} &=& \Omega^{AB}.
\eea

The $SO(6)$ invariant half-gamma matrices with properties
\bea
\frac{1}{2} \epsilon^{ABCD} \Sigma^{M}_{CD} &=& \Sigma^{MAB},\label{cond1}\\
\Sigma^{*M}_{AB} &=& -\Sigma^{MAB}.\label{cond2}
\eea
are not q-tensors. But we can express them in terms of q-tensors $\Gamma^m_{AB}$ and $\Omega_{AB}$ if we reduce $SO(6)$ to $SO(5)$ where we decompose the vector index as $M=(m,6)$,
\bea
\Sigma^M_{AB} &=& (\Gamma^m_{AB}, i \Omega_{AB}),\cr
\Sigma^{MAB} &=& (-\Gamma^{mAB}, i \Omega^{AB}).
\eea
We then find
\bea
\Gamma^{mAB} &=& \Gamma^{*m}_{AB}.
\eea
The $\N=6$ supersymmetry variation
\bea
\delta Z^A_{\qa} &=& i\bar{\epsilon}^{AB} \psi_{B\qa},\cr
\delta (Z^A_{\qa})^* &=& -i\bar{\epsilon}_{AB} (\psi_{B\qa})^*
\eea
where the supersymmetry parameters are given by
\bea
\epsilon_{AB} &=& \epsilon^M \Sigma^M_{AB},\cr
\epsilon^{AB} &=& \epsilon^M \Sigma^{MAB}
\eea
is decomposed as
\ben
\delta Z^A_{\qa} &=& -i\bar{\epsilon}^m \Gamma^{mAB} \psi_{Ba} - \bar{\epsilon} \Omega^{AB} \psi_{B\qa},\label{nc}\\
\delta (Z^A_{\qa})^* &=& -i\bar{\epsilon}^m \Gamma^m_{AB} (\psi_{B\qa})^* + \bar{\epsilon} \Omega_{AB} (\psi_{B\qa})^*\label{cc}
\een
It is now interesting to see what happens to this if we impose the reality conditions on the matter fields as specified above, that is $(Z^A_{\qa})^* = -Z_A^{\qa}$ and $(\psi_{A\qa})^* = \psi^{A\qa}$. We then get from (\ref{cc}),
\bea
\delta Z_A^{\qa} &=& i\bar{\epsilon}^m \Gamma^m_{AB} \psi^{B\qa} - \bar{\epsilon} \Omega_{AB} \psi^{B\qa}
\eea
Treating all quantities in (\ref{nc}) as q-tensors we instead get from (\ref{nc})
\bea
\delta Z_A^{\qa} &=& i\bar{\epsilon}^m \Gamma^m_{AB} \psi^{B\qa} + \bar{\epsilon} \Omega_{AB} \psi^{B\qa}.
\eea
Hence the reality conditions are compatible with $\N=5$ supersymmetry, but in general not with $\N=6$ supersymmetry.

\subsection{The q- and h-algebra formulations of $\N=5$ theories}
We have seen that the q-brackets can be expressed in terms of h-brackets. But the converse is not true. We can not write $\N=6$ ABJM theory in terms of q-brackets. However if we impose certain reality conditions on the matter fields this becomes possible. The $\N=5$ supersymmetric Lagrangian can be expressed entirely in terms of q-brackets as
\bea
\L &=& -D_{\mu}Z_{A\qa} D^{\mu} Z^{A\qa} + i\bar{\psi}_{A\qa} \gamma^{\mu} D_{\mu} \psi^{A\qa} - V(Z)\cr
&& + i\(-(\Omega^{-1})^{AC}(\Omega^{-1})^{BD}+2(\Omega^{-1})^{AD}(\Omega^{-1})^{BC}\)\bar{\psi}_{A\qa}Z_{B\qb}\psi_{C\qc}Z_{D\qd}f^{\qb\qc\qd\qa}\cr
&& + \L_{CS}
\eea
where the sextic potential is given by
\bea
V(Z) &=& \frac{2}{3}\(|\{Z_A,Z_B,Z_C\}|^2 + |\{Z_A,Z_B,Z_C\}\(\Omega^{-1}\)^{AC}|^2\)
\eea
where $|X|^2 = h(X,X)$ and the Chern-Simons term is given by 
\bea
\L_{CS} &=& \frac{1}{2} \(f^{\qa\qb\qc\qd} A_{\qc\qb} d A_{\qd\qa} + \frac{1}{3} f^{\qe\qa\qb}{}_{\qg} f^{\qg\qc\qd\qf} A_{\qa\qb}A_{\qc\qd}A_{\qe\qf}\).
\eea
The $\N=5$ supersymmetry variations are
\bea
\delta Z_{A\qa} &=& -i \bar{\epsilon}_A{}^B \psi_{B\qa},\cr
\delta \psi_{A\qa} &=& -\gamma^{\mu} \epsilon_A{}^B D_{\mu} Z_{B\qa}\cr
&& + \(\frac{2}{3}\epsilon_A{}^B \{Z_C, Z_B, Z_D\} - \frac{4}{3} \epsilon_D{}^B \{Z_C, Z_B, Z_A \}\)_{\qa} (\Omega^{-1})^{CD},\cr
\delta \bar{A}_{\mu}{}^{\qb}{}_{\qa} &=& 2i \bar{\epsilon}^{AB} \gamma_{\mu} \psi_{A\qc} Z_{B\qd} f^{\qb\qc\qd}{}_{\qa}
\eea
In terms of matrices the $\N=5$ supersymmetry variations become
\bea
\delta Z_A &=& -i \bar{\epsilon}_A{}^B \psi_B,\cr
\delta \psi_A &=& -\gamma^{\mu} \epsilon_A{}^B D_{\mu} Z^B\cr
&&- \frac{2\lambda}{3}\epsilon^C{}_{A} \(Z_{[C} \t{Z}^B Z_{B]} + \frac{1}{2}\(Z^B \t{Z}^C Z_B - Z_B \t{Z}^C Z^B\)\) \cr
&&+ \frac{4\lambda}{3}\epsilon^B{}_C \(Z_{[A} \t{Z}^C Z_{B]} + Z^C \t{Z}_A Z_B\),\cr
\delta A^R_{\mu} &=& i\lambda \bar{\epsilon}^{AB}\gamma_{\mu}\(\t Z_B \psi_A + \t \psi_A Z_B\),\cr
\delta A^L_{\mu} &=& i\lambda \bar{\epsilon}^{AB}\gamma_{\mu}\(\psi_A \t Z_B + Z_B \t \psi_A\). 
\eea
Specializing to gauge groups $Sp(N)_K\times O(M)_{-K}$ we have essentially reproduced the results in \cite{Hosomichi:2008jb}.

The transition to h-algebra is made by splitting the symmetric q-bracket into antisymmetric h-brackets according to Eq (\ref{split}). We begin by rewriting the supersymmetry variations. The $\N=5$ supersymmetry variation of the scalar field can be written in the form
\bea
\delta Z^A_{\qa} &=& -i\bar{\epsilon}^{AB} \psi_{B\qb}.
\eea
The supersymmetry variation of the fermion becomes
\bea
\delta \psi_{A\qa} &=& \gamma^{\mu} \epsilon_{AB} D_{\mu} Z^B_{\qa} \cr
&& +\(\frac{2}{3}\epsilon_{AB} Z^B_{\qb} Z^C_{\qc} Z_{C\qc} + \frac{4}{3} \epsilon_{BC} Z^B_{\qb} Z^C_{\qc} Z_{A\qd}\) g^{\qb\qc\qd}{}_{\qa} \cr
&& + \(-\frac{2}{3} \epsilon_{AB} Z^C_{\qb} Z_{C\qc} Z^B_{\qd} + \frac{4}{3} \epsilon_{BC} Z^B_{\qb} Z_{A\qc} Z^C_{\qd}\) g^{\qb\qc\qd}{}_{\qa}
\eea
The second line is not of the form of the $\N=6$ supersymmetry variation. But we can bring the second line into this form as follows,
\bea
&&\(-\frac{2}{3} \epsilon_{AB} Z^C_{\qb} Z_{C\qc} Z^B_{\qd} + \frac{4}{3} \epsilon_{BC} Z^B_{\qb} Z_{A\qc} Z^C_{\qd}\) g^{\qb\qc\qd}{}_{\qa}\cr
&=& \frac{2}{3}  \(\Omega_{AH} \Omega_{EF} - 2\Omega_{AF} \Omega_{EH}\)\epsilon^{HG} Z^E_{\qb} Z^F_{\qc} Z_G^{\qd} g^{\qb\qc}{}_{\qd\qa}
\eea
We then use 
\bea
\epsilon^{HG} &=& \frac{1}{2} \epsilon^{HGBC} \epsilon_{BC}
\eea
and the second line becomes
\bea
\(-\frac{2}{3} \epsilon_{BC} Z^B_{\qb} Z^C_{\qc} Z_{A}^{\qd} -\frac{4}{3} \epsilon_{AB} Z^B_{\qb} Z^C_{\qc} Z_C^{\qd}\) g^{\qb\qc}{}_{\qd\qa}
\eea
Adding this to the first line we have
\bea
\delta \psi_{A\qa} &=&  \gamma^{\mu} \epsilon_{AB} D_{\mu} Z^B_{\qa} - \(\epsilon_{AB} Z^B_{\qb} Z^C_{\qc} Z_C^{\qd} - \epsilon_{BC} Z^B_{\qb} Z^C_{\qc} Z_A^{\qd}\) (2g)^{\qb\qc}{}_{\qd\qa}
\eea
For the gauge field we get get
\bea
\delta \bar{A}_{\mu}{}^{\qb}{}_{\qa} &=& \(-i \bar{\epsilon}^{AB}\gamma_{\mu} \psi_{A\qc} Z_{B}^{\qd} + i \bar{\epsilon}_{AB}\gamma_{\mu} Z^A_{\qc} \psi^{B\qd} \) (2g)^{\qb\qc}{}_{\qd\qa}.
\eea

Let us next turn to the action. For the Yukawa couplings we get
\bea
&& i\((\Omega^{-1})^{AC} (\Omega^{-1})^{BD} + 2(\Omega^{-1})^{AD} (\Omega^{-1})^{BC})\)\bar{\psi}_{A\qa} Z_{B\qb} \psi_{C\qc} Z_{D\qd} \(g^{\qb\qc\qd\qa} + g^{\qb\qd\qc\qa}\)\cr
&=& -i\(\bar{\psi}^{C\qa} \psi_{C\qb} Z^D_{\qc} Z_D^{\qd} + 2 \bar{\psi}^{D\qa} \psi_{C\qb} Z^C_{\qc} Z_D^{\qd}\) g^{\qb\qc}{}_{\qd\qa}\cr
&& + i \epsilon_{ABCD} \bar{\psi}^{A\qa} Z^B_{\qb} Z^C_{\qc} \psi^{D\qd} g^{\qb\qc}{}_{\qd\qa}.
\eea
For the sextic potential we get two terms
\bea
\tr\(\{Z^A,Z^B,\t Z^C\}\{Z_A,Z_B,\t Z_C\}\) &=& -4\tr\(Z^A Z_C Z^B Z_A Z^C Z_B + Z_A Z^A Z_C Z^B Z_B Z^C\),\cr
\tr\(\{Z^A,Z^B,\t Z_A\}\{Z_C,Z_B,\t Z^C\}\) &=& 2\tr\(Z^A Z_A Z^B Z_C Z^C Z_B - Z^A Z_A Z^B Z_B Z^C Z_C\)
\eea
For the Chern-Simons terms we get
\bea
f^{\qa\qb\qc\qd} A_{\qc\qb} dA_{\qd\qa} &=& 2g^{\qa\qb}{}_{\qc\qd} A^{\qc}{}_{\qb} dA^{\qd}{}_{\qa},\cr
f^{\qe\qa\qb}{}_{\qg} f^{\qg\qc\qd\qf} A_{\qa\qb}A_{\qc\qd}A_{\qe\qf} &=& 4g^{\qe\qa}{}_{\qb\qg} g^{\qg\qc}{}_{\qd\qf} A_{\qa}{}^{\qb}A_{\qc}{}^{\qd}A_{\qe}{}^{\qf}
\eea
We also note that, by the reality conditions,
\bea
\tr\(Z^A Z_A Z^B Z_B Z^C Z_C\) &=& -\tr\(Z_A Z^A Z_B Z^B Z_C Z^C\),\cr
\epsilon_{ABCD} \bar{\psi}^{A\qa} Z^B_{\qb} Z^C_{\qc} \psi^{D\qd} g^{\qb\qc}{}_{\qd\qa} &=& \epsilon^{ABCD} \bar{\psi}_{A\qa} Z_B^{\qb} Z_C^{\qc} \psi_{D\qd} g^{\qd\qa}{}_{\qb\qc}.
\eea

The $\N=5$ supersymmetry variations and the $\N=5$ Lagrangian are brought into the same form as $\N=6$ ABJM theory once we formulate everything in terms of h-algebra. Naively one may think \cite{Bagger:2010zq} that not every $\N=5$ theory can be derived from $\N=6$ ABJM theory by imposing reality conditions while keeping the h-algebra structure unchanged since that surely gives rise to gauge groups which are not the ones one would expect to see in the purely $\N=5$ theories (and which are not just restriction of $\N=6$). The resolution is that we have two equivalent formulations of $\N=5$ theory. One is the q-algebra formulation, the other is the h-algebra formulation. Which gauge group we have, depends on which formulation we use. The $\N=6$ theories have only the h-algebra formulation. That is how we see that $\N=6$ theories constitute  a proper subset of the $\N=5$ theories.

The most general form for the $\N=5$ supersymmetry variations was obtained in \cite{Bagger:2010zq}. In this reference it is not clearly stated, but it can be immediately seen by rising and lowering indices by symplectic forms, that these most general $\N=5$ supersymmetry variations can be expressed as (in our conventions)
\bea
\delta Z^A_{\qa} &=& -i\bar{\epsilon}^{AB} \psi_{B\qb},\cr
\delta \psi_{A\qa} &=&  \gamma^{\mu} \epsilon_{AB} D_{\mu} Z^B_{\qa} - \(\epsilon_{AB} Z^B_{\qb} Z^C_{\qc} Z_C^{\qd} - \epsilon_{BC} Z^B_{\qb} Z^C_{\qc} Z_A^{\qd}\) (2g)^{\qb\qc}{}_{\qd\qa},\cr
\delta \bar{A}_{\mu}{}^{\qb}{}_{\qa} &=& \(-i \bar{\epsilon}^{AB}\gamma_{\mu} \psi_{A\qc} Z_{B}^{\qd} + i \bar{\epsilon}_{AB}\gamma_{\mu} Z^A_{\qc} \psi^{B\qd} \) (2g)^{\qb\qc}{}_{\qd\qa}.
\eea
where the structure constants obey
\ben
(2g)^{\qb\qc}{}_{\qd\qa} &=& -(2g)^{\qc\qb}{}_{\qd\qa}\label{antisymmetri}
\een
These variations are exactly of the form of the $\N=6$ supersymmetry variations. The difference is that we have in addition reality conditions on the matter fields, which are absent in the $\N=6$ theories. Closure of these supersymmetry variation requires the structure constants obey the hermitian fundamental identity. In \cite{Bagger:2010zq}, in an attempt to find more general theories than those which have $\N=6$ supersymmetry, these structure constants were antisymmetrized by hand and written as $g^{\qb\qc}{}_{\qd\qa} - g^{\qc\qb}{}_{\qd\qa}$ instead. Since these  occur everywhere being antisymmetrized, we should instead be allowed to assume that they are already antisymmetric as specified by Eq (\ref{antisymmetri}). Later on the by hand performed antisymmetry of the structure constants is forgotten. But this then gives us too many possible gauge groups, most of which we can not distinguish among in the $\N=5$ theories. We can only distinguish those gauge groups that correspond to antisymmetric three-algebra structure constants. 

The goal in \cite{Bagger:2010zq} was to find the $\N=5$ theories from a three-algebraic approach, which were found using an embedding tensor formalism in \cite{Bergshoeff:2008bh}. Here we have a linear constraint which we will interpret in terms of a constraint on the q-algebra structure constants
\ben
f^{(\qa\qb\qc\qd)} &=& 0\label{emb}
\een
which is a sum of $24$ terms. The other constraint found on the embedding tensor now corresponds to the q-algebra fundamental identity. 

First we check that one class of solutions to the linear constraint is indeed given by symmetric q-algebra structure constants subject to 
\ben
f^{\qa\qb\qc\qd} &=& f^{\qa\qc\qb\qd},\label{sym1}\\
f^{\qa\qb\qc\qd} &=& f^{\qc\qd\qa\qb},\label{sym2}
\een
which are such that they can be split into
\ben
f^{\qa\qb\qc\qd} &=& g^{\qa\qb\qc\qd} + g^{\qa\qc\qb\qd}.\label{splitt}
\een
We note that the embedding tensor always obeys (\ref{sym2}). The only extra assumption then is that we also have the symmetry (\ref{sym1}). With this assumption we get
\bea
f^{(\qa\qb\qc\qd)} &\propto & f^{\qa\qb\qc\qd} + f^{\qa\qd\qb\qc} + f^{\qa\qc\qd\qb}
\eea
and we need to check that this vanishes when inserting (\ref{splitt}). Doing the insertion and rearranging the terms, we get
\bea
f^{(\qa\qb\qc\qd)} &\propto & g^{\qa\qb\qc\qd} + g^{\qa\qb\qd\qc} + g^{\qa\qc\qb\qd} + g^{\qa\qc\qd\qb} + g^{\qa\qd\qb\qc} + g^{\qa\qd\qc\qb} 
\eea
Now each pair vanishes by antisymmetry
\bea
g^{\qa\qb\qc\qd} &=& - g^{\qb\qa\qc\qd},\cr
g^{\qa\qb\qc\qd} &=& g^{\qc\qd\qa\qb}.
\eea
That establishes that our solutions are not in contradiction with the embedding tensor formalism.

\subsection{Enhanced $\N=6$ supersymmetry by restriction}
A trivial enhancement of supersymmetry arises if we explicitly solve and eliminate the reality conditions. In that process we also eliminate the symplectic form $\omega_{\qa\qb}$. As we then have no symplectic form, we must necessarily use h-algebra and q-algebra is no longer an option. Then we have reduced to $\N=6$ ABJM theory. 

In order to eliminate\footnote{What from $\N=6$ theory point of view appears like eliminating the symplectic form, does from the $\N=5$ theory point of view look like we just make some choice of the symplectic form. We shall have no symplectic form in $\N=6$ theory.} the symplectic form, we split the q-algebra index as
\bea
\qa &=& (\aa,\pm)
\eea
and define the symplectic form in the $\N=5$ theory as
\bea
\omega_{\aa+,\bb-} &=& -\delta_{\aa}^{\bb}
\eea
and being antisymmetric in the indices $\pm$ and hence symmetric in $\aa,\bb$. The inverse is given by 
\bea
(\omega^{-1})^{\aa+,\bb-} &=& \delta^{\bb}_{\aa}.
\eea
We can solve the reality conditions as 
\bea
Z_{A\aa+} &=& Z_A^{\aa},\cr
Z_{A\aa-} &=& \Omega_{AB} Z^B_{\aa},
\eea
\bea
\psi_{A\aa+} &=& \Omega_{AB} \psi^B_{\aa},\cr
\psi_{A\aa-} &=& \psi_{A}^{\aa} 
\eea
and 
\bea
\t A_{\mu}{}^{\bb+}{}_{\aa+} &=& \t A_{\mu}{}^{\bb}{}_{\aa},\cr
\t A_{\mu}{}^{\bb-}{}_{\aa-} &=& -\t A_{\mu}{}^{\aa}{}_{\bb}.
\eea
with all other components vanishing. Here
\bea
Z^{*A}_{\aa} &=& Z_A^{\aa},\cr
\psi_{A\aa}^* &=& \psi^{A\aa}.
\eea
If we now restrict the gauge group such that  
\bea
(2g)^{\bb+\cc+}{}_{\dd+\aa+} &=& f^{\bb\cc}{}_{\dd\aa},\cr
(2g)^{\bb-\cc-}{}_{\dd-\aa-} &=& f^{\bb\cc}{}_{\dd\aa}
\eea
and all other components vanish, then we find that these complex components get identified as the matter fields of $\N=6$ ABJM theory. The action recieves an overall factor of $2$. Since for instance $\t A_{\mu}{}^{\bb+}{}_{\aa-} = 0$ any mixed term such as $(2g)^{++}{}_{++}(2g)^{--}{}_{--}A^-{}_+ A^+{}_+ A^-{}_-$ will vanish. All that happens is that we get an overall rescaling of the action by this factor of $2$. If we started by writing the $\N=6$ theory in the $\N=5$ form with real matter fields and with Chern-Simons level $K$, we find the Chern-Simons level $2K$ when the same theory is expressed in terms of the $\N=6$ ABJM action.

\subsection{Enhanced $U(4)$ R-symmetry by monopoles} 
If we keep the reality conditions then we are in the regime of $\N=5$ theories that are not $\N=6$. However it can still happen that we have enhanced R symmetry if there is a monopole field which changes the reality condition into the opposite one. This is easiest seen by considering the R symmetry. The $Sp(2)\simeq SO(5)$ R-symmetry rotates the four complex scalar fields as
\bea
\delta Z_{A\qa} &=& \omega_A{}^B Z_{B\qa}
\eea
where $\omega_A{}^B$ leaves the $Sp(2)$ invariant form $\Omega_{AB}$ invariant,
\ben
\omega_{AB} - \omega_{BA} &=& 0\label{sp}
\een
where $\omega_{AB} = \Omega_{BC} \omega_A{}^C$. This leaves us with $10$ parameters, which is the number of generators of $Sp(2) \simeq SO(5)$. To enhance the R-symmetry to $U(4)$ we need $6$ more generators and we need a monopole field $\M_{\qa}{}^{\qb}$ which can be used to dress the scalar field
\bea
\t Z_{A\qa} &=& \M_{\qa}{}^{\qb} Z_{A\qb}
\eea
which changes the reality condition as
\bea
Z_{A\qa}^* &=& Z^{A\qa},\cr
\t Z_{A\qa}^* &=& -\t Z^{A\qa}.
\eea
Then $U(4)$ can act infinitesimally as
\bea
\(\begin{array}{c}
\delta Z_{A\qa}\\
\delta \t Z^A_{\qa}
\end{array}\) &=& 
\(\begin{array}{cc}
\omega_A{}^B & \eta_{AB}\\
\eta^{AB} & -\omega_B{}^A
\end{array}\)
\(\begin{array}{c}
Z_{A\qa}\\
\t Z^A_{\qa}
\end{array}\)
\eea
Indeed this matrix is anti-hermitian if we assign 
\bea
\omega^*_A{}^B &=& -\omega_B{}^A,\cr
\eta^*_{AB} &=& -\eta^{AB}.
\eea
All quantities now behave consistently like q-tensors and we have enhanced $U(4)$ symmetry.

\subsection{Properties of the monopole field}
What we need for supersymmetry enhancement is a local monopole field $\M_{\qa}{}^{\qb}$ that can be used to convert a matter field $X_{A\qa}$ to a dressed matter field $\t X_{A\qa}$,
\bea
\t X_{A\qa} &=& \M_{\qa}{}^{\qb} X_{A\qb}
\eea
with opposite reality condition, so that we have
\bea
\t X_{A\qa}^* &=& - \t X^{A\qa},\cr
X_{A\qa}^* &=& X^{A\qa}.
\eea
From this we derive the following reality condition on the monopole field,
\ben
\M^*_{\qa\qb} &=& -\M^{\qa\qb}.\label{re}
\een
We have used $\omega_{ab}$ to put both indices down. When doing this one should be careful with the complex conjugation since that brings in a sign coming from $\omega_{\qa\qb}^* = -(\omega^{-1})^{\qa\qb}$. Now this reality condition for the monopole field, should be contrasted with the corresponding reality condition for the symplectic form
\bea
\omega^*_{\qa\qb} &=& \omega^{\qa\qb}.
\eea
Clearly we can never identify the monopole field with the symplectic form. They satisfy opposite reality conditions. 

A priori we can imagine the monopole being either symmetric or antisymmetric,
\bea
\M_{\qa\qb} &=& s \M_{\qb\qa}
\eea
As it turns out we then get
\bea
(\M^{-1})_{\qa}{}^{\qb} &=& -s \M_{\qa}{}^{\qb}
\eea
The monopole which can enhance R symmetry to $U(4)$ must be symmetric as one can see by making a $U(4)$ transformation of the kinetic terms $DZ_{A\qa} DZ^{A\qa}$. We can construct the monopole field using Wilson lines $W_{\qa}{}^{\qa'}$ with the index $\qa'$ at infinity. We define the monopole field as
\bea
\M_{\qa\qb} &=& \delta_{\qa'\qb'} W_{\qa}{}^{\qa'} W_{\qb}{}^{\qb'}
\eea
which is symmetric. The line is invisible and extends to infinity where we have the primed index $\qa'$. Complex conjugation reverses the orientation of the Wilson line,
\bea
W^*_{\qa}{}^{\qa'} &=& (W^{-1})_{\qa'}{}^{\qa}.
\eea
Then 
\ben
\M^*_{\qa\qb} &=& \delta^{\qa'\qb'} (W^{-1})_{\qa'}{}^{\qa} (W^{-1})_{\qb'}{}^{\qb}\label{complexx}
\een
and we find that 
\bea
\M^*_{\qa\qb} \M_{\qb\qc} &=& \delta_{\qc}^{\qa}.
\eea
which can also be expressed as
\ben
(\M^{-1})_{\qa}{}^{\qb} &=& \M_{\qa}{}^{\qb}\label{sq}
\een
or, equivalently,
\bea
\t{\t Z}_{A\qa} &=& Z_{A\qa}.
\eea
From (\ref{sq}) we get 
\bea
\M^{\qa\qb} \M_{\qb\qc} = (\omega^{-1})^{\qa\qe} \omega_{\qc\qf} \delta^{\qf}_{\qe} = -\delta^{\qa}_{\qc}
\eea
By noting (\ref{complexx}) we see that the reality condition (\ref{re}) that is required of the monopole field in order to enhance supersymmetry, is now realized by construction.

Let us denote the inner product as $\<X,Y\> = (\omega^{-1})^{\qa\qb} X_{\qa} Y_{\qb} = X_{\qa} Y^{\qa}$ reflecting the fact the this inner product can be viewed as both symplectic as well as hermitian (as we impose reality conditions on the matter fields). The monopole field has the property 
\bea
\<X,\t Y\> &=& \<\t X,Y\>
\eea
Also we have (the notation will be clarified below)
\bea
\t X \t Y \t Z &=& \t {XYZ}.
\eea
This can be shown by the following computations,
\bea
(\omega^{-1})^{\qa\qb} X_{\qa} \M_{\qb}{}^{\qc} Y_{\qc} = \M^{\qa\qc} X_{\qa} Y_{\qc} = -\M^{\qc\qa} X_{\qa} Y_{\qc} = (\omega^{-1})^{\qa\qb} \M_{\qa}{}^{\qc} X_{\qc} Y_{\qb}
\eea
and
\bea
\t X \t Y \t Z &\equiv & \M_{\qb}{}^{\qe} \M_{\qc}{}^{\qf} \M_{\qd}{}^{\qg} X_{\qe} Y_{\qf} Z_{\qh} g^{\qb\qc\qd}{}_{\qa} = \M_{\qa}{}^{\qe} X_{\qb} Y_{\qc} Z_{\qd} g^{\qb\qc\qd}{}_{\qe} \equiv \t {XYZ}. 
\eea
where we have used the invariance of the structure constants. 

The monopole field is a local composite field which transforms covariantly according to its placings of gauge indices. This means that
\ben
\t {D_{\mu} T} &=& D_{\mu} \t T.\label{monopole}
\een
where we may express the gauge covariant derivative in terms of indices as
\bea
D_{\mu} T_{\qa_1\qa_2...}^{\qb_1\qb_2..} &=& \partial_{\mu} T_{\qa_1\qa_2...}^{\qb_1\qb_2..} + \(\bar{A}_{\mu \qa_1}{}^{\qc} T^{\qb_1\qb_2..}_{\qc\qa_2..}+..\) - \(T_{\qa_1\qa_2..}^{c\qb_2..}\bar{A}_{\mu\qc}{}^{\qb_1} +..\)
\eea
Gauge covariance (\ref{monopole}) implies that  
\bea
D_{\mu} \M &=& 0.
\eea
Using the above form of the covariant derivative, and noting that the index structure on the monopole field is $\M_{\qa}{}^{\qb}$, we can express this condition as
\bea
\partial_{\mu} \M + [\bar{A}_{\mu},\M] &=& 0.
\eea
If $\Delta$ denotes an infinitesimal symmetry variation of the theory, then we must also have that 
\bea
\Delta (D_{\mu} \M) &=& 0.\label{qqq}
\eea
If the gauge field is put on-shell then the condition $D_{\mu} \M = 0$ is a constraint on the matter fields. If on the other hand we do not put the gauge field on-shell then we do not have a constraint on the matter fields. Then this is rather a property of the Wilson line. The constraint on the matter fields is a quantum effect obtained by integrating out the gauge field. The constraints are not imposed in the classical action and they do not change the field content of the classical theory. These constraints shall be closed under any symmetry of the theory. In particular they form multiplets of $Sp(2)$ R symmetry and $\N=5$ supersymmetry. We can rewrite (\ref{qqq}) as 
\bea
\Delta \bar{A}_{\mu} &=& \M (\Delta \bar{A}_{\mu}) \M^{-1} - D_{\mu}\(\M \Delta \M^{-1}\)
\eea
We will in all our equations discard the term which is an infinitesimal gauge variation with gauge parameter $\M \Delta \M^{-1}$. This will be consistent by the same argument as in {\cite{Gustavsson:2009pm}. When then have the constraint
\bea
\Delta \bar{A}_{\mu} &=& \M (\Delta \bar{A}_{\mu}) \M^{-1}.
\eea

\subsubsection{Two constraints}
In order to understand supersymmetry enhancement from $\N=5$ to $\N=6$, we choose to formulate the $\N=5$ theory in h-algebra language. We can then shorten our expressions, as it is always clear how we shall put the three-algebra indices in the h-algebra formulation. We shall put the indices according to 
\bea
X_{\qb} Y_{\qc} Z^{\qd} g^{\qb\qc}{}_{\qd\qa} &\equiv (X Y Z)_{\qa},\cr
Y_{\qc} Z^{\qd} g^{\qb\qc}{}_{\qd\qa} &\equiv (Y Z)^{\qb}{}_{\qa}.
\eea
The right-hand sides will be our short-hand expressions for these index structures. No other index structures will appear so this notation is not ambiguous. However, while we thus stick to the h-algebra for the gauge indices, we do not necessarily stick to full $\N=6$ form on the supersymmetry variations. We will use the other symplectic form $\Omega_{AB}$ to rise and lower R-symmetry indices freely.

From the $\N=5$ supersymmetry variation of the gauge field,
\bea
\Delta \bar{A}_{\mu} &=& i \bar{\epsilon}^m \gamma_{\mu} \Gamma^{mAB} \(\psi_A Z_B + Z_B \psi_A\)
\eea
we obtain the constraint
\ben
\Gamma^{mAB}\(\psi_A Z_B + \t Z_B \t \psi_A\) + \Gamma^{mAB}\(Z_B \psi_A + \t \psi_A \t Z_B\) &=& 0\label{first}
\een
We have split this into two terms, each of them corresponding each term in $\delta \bar{A}$ respectively. 

This motivates us to consider three distinct quantities that we define as follows,
\bea
M^m &=& \Gamma^{mAB} \(\psi_A Z_B + \t Z_B \t \psi_A\),\cr
N^{m\pm} &=& i\Gamma^{mAB} \(\psi_A Z_B \pm \t \psi_A \t Z_B\).
\eea
These are subject to 
\bea
\t M^m &=& -M^m,\cr
\t N^{m \pm} &=& \pm N^{m \pm}
\eea
and 
\bea
M^{*m} &=& M^m,\cr
N^{*m\pm} &=& \mp N^{m\pm}.
\eea 
We see that $N^{m-}$ and $M^m$ are subject to the same projection properties. We now claim that the right constraints are given by
\ben
M^m &=& 0,\label{ett}\\
N^{m-} &=& 0\label{tva}
\een
It is not reasonable to have both the constraints $N^{m+} = 0$ and $N^{m-} = 0$. Combining these we would then also have a constraint $\Gamma^{mAB} \psi_A Z_B = 0$ that does not involve a monopole operator at all. Such a constraint is unacceptable since it would affect the $\N=5$ theory itself. The $\N=5$ theory shall be untouched, and we just want to check that it under certain circumstances receives enhanced $\N=6$ supersymmetry. We shall not modify the $\N=5$ theory at all and therefore we can impossibly have both of the constraints $N^{m\pm} = 0$. 

If we agree on that $M^m = 0$ shall be one set of constraints, then it is also natural that also $N^{m-} = 0$ shall be another set of constraints since these obey the same projection conditions. 

Even though our motivation for these constraints is somewhat vague, we will see that many things will fit together once we agree on that these constraints are the correct ones. 

\subsubsection{Further constraints}
The constraint $M^m = 0$ can be derived from the constraint
\bea
Z^A Z_A - \t Z^A \t Z_A &=& 0.
\eea
By making an $\N=5$ supersymmetry variation we get
\bea
\Gamma^{mAB}\(\psi_A Z_B + \t Z_B \t \psi_A\) - \Gamma^{mAB}\(Z_B \psi_A + \t \psi_A \t Z_B\) &=& 0
\eea
with a relative minus sign between the terms. But according to the results in the previous subsection, both of these terms vanish being constraints. 

We will now show that by making an $\N=5$ supersymmetry variation of (\ref{ett}) we get the constraint
\ben
X Z_A Y + \t Y Z_A \t X &=& 0\label{etta}
\een
and by making an $\N=5$ supersymmetry variation of (\ref{tva}) we get the constraint 
\ben
X Z_A Y - \t X Z_A \t Y &=& 0\label{tvaa}
\een
Here $X$ and $Y$ can be any matter field, either bosonic or fermionic. Let us just consider the latter case since the former case is done along the same lines. We thus start by 
\bea
\Gamma^{mAB} \([X,\psi_A;Z^B]-[X,\t \psi_A;\t Z^B]\) &=& 0
\eea
and make another $\N=5$ variation. We then consider the inner product with some other matter field, let us denote it as $Y$. Then we have
\bea
\<[X,[Z^C,Z^D;Z^D];Z^B],Y\> - \<[X,[\t Z^C,\t Z^D;\t Z^D];\t Z^B],Y\> &=& 0
\eea
We use the invariance of the inner product to get
\bea
\<[Z^C,Z^D;Z^D],[Y,Z^B,X]\> - \<\t{[Z^C,Z^D;Z^D]},[Y,\t Z^B,X]\> &=& 0
\eea
We use the property of the monopole field to get
\bea
\<[Z^C,Z^D;Z^D],[Y,Z^B,X]\> + s\<[Z^C,Z^D;Z^D],\t {[Y,\t Z^B,X]}\> &=& 0
\eea
and again another property of the monopole (that it squares to $-s$) and get
\bea
\<[Z^C,Z^D;Z^D],[Y,Z^B,X]\> - \<[Z^C,Z^D;Z^D],[\t Y,Z^B,\t X]\> &=& 0.
\eea
As it should, the form of this constraint does not depend on $s$. Since $Z^C$ can be varied independently of the other components, we conclude that we have the constraint (\ref{tvaa}). The derivation of (\ref{etta}) is done in a similar way.

\subsection{Enhanced $\N=6$ supersymmetry}
We need to find one extra $\N=1$ supersymmetry to get from $\N=5$ to $\N=6$. The form of the $\N=1$ supersymmetry variation is given by
\bea
\delta Z_A &=& - \bar{\epsilon} \Omega_{AB} \t \psi^B,\cr
\delta \psi_A &=& i\gamma^{\mu} \epsilon \Omega_{AB} D_{\mu} \t Z^B - i\epsilon \(\Omega_{AB} \t Z^B Z^C Z_C + \Omega_{BC} Z^B Z^C \t Z_A\),\cr
\delta \bar{A}_{\mu} &=& \bar{\epsilon} \gamma_{\mu} \Omega^{AB} \(\psi_A \t Z_B + \t Z_B \psi_A\).
\eea
We must have an odd number of tildes (either one or three) in the right-hand side for the variations to become compatible with the reality conditions. But for the three-brackets there is no unique or most natural way to distribute one or three tildes. The resolution will be that the various possibilies one has are all equivalent thanks to constraints on the matter fields. 

Using the constraint (\ref{tvaa}) we get instance
\bea
\t Z^B Z^C Z_A &=& Z^B Z^C \t Z_A
\eea
and by antisymmetry this is also equal to
\bea
-Z^C \t Z^B Z_A.
\eea
This shows that it does not matter on which $Z$ we put a single tilde in the term $\Omega_{BC} \t Z^B Z^C Z_A$. Similarly 
\bea
\t Z^B Z^C Z_C &=& Z^B Z^C \t Z_C
\eea
and by antisymmetry this is also equal to 
\bea
-Z^C Z^B \t Z_C &=& -\t Z^C Z^B Z_C\cr
&=& Z^B \t Z^C Z_C
\eea
and so we see that also for this term it does not matter on which $Z$ we put the tilde. But we can do more, since we have
\bea
\t Z^B Z^C Z_C &=& \t Z^B \t Z^C \t Z_C
\eea
so we also see that we can also put three tildes on this term and again get the same result. Similarly we have 
\bea
\t Z^B Z^C Z_A &=& - Z^C \t Z^B Z_A \cr
&=& - \t Z^C \t Z^B \t Z_A \cr
&=& \t Z^B \t Z^C \t Z_A.
\eea

\subsubsection{Commuting $\N=1$ with $\N=1$}
When we commute two $\N=1$ variations, we get no term proportional to $\bar{\epsilon}\eta$ since $\bar{\epsilon} \eta - \bar{\eta} \epsilon = 0$. This is true irrespectively of the distribution of the tildes. In particular then, we get no gauge variation term. 

For the scalars we get
\bea
[\delta^1_{\eta},\delta^1_{\epsilon}] Z^A &=& -2i \bar{\epsilon}\gamma^{\mu}\eta D_{\mu} Z^A
\eea

For the fermions we get
\bea
[\delta^1_{\eta},\delta^1_{\epsilon}] \psi_A &=& -2i \bar{\epsilon}\gamma^{\mu}\eta D_{\mu} \psi_A + \bar{\epsilon} \gamma^{\lambda} \gamma_{\lambda} \eta E_{A}
\eea
where $E_A$ is the $\N=5$ equation of motion. To see this we need to use constraints in the form
\bea
\t Z^B \psi_C \t Z_D = \t Z^B Z^C \t \psi^D = Z^B Z^C \psi^D,\cr
\t Z^B \psi_C \t Z_D = \t Z^B \t \psi^C Z_D.
\eea 

For the gauge field variation we must be more careful with the distribution of the tilde (or the monopole field). Only if we knew that we had $\N=1$ supersymmetry, we could use that $\delta \bar{A} = \t{\delta \bar{A}}$. Here we instead choose the following particular distribution of one tilde in the gauge field variation,
\bea
\delta \bar{A}_{\mu} &=& \bar{\epsilon} \gamma_{\mu} \Omega^{AB} \(\psi_A \t Z_B + \t Z_B \psi_A\)
\eea
and try to close all these variations on-shell. We find that the $\N=1$ variations close among themselves on the gauge field equation of motion 
\bea
F_{\mu\nu} &=& \epsilon_{\mu\nu\lambda} \t Z^A D^{\lambda} \t Z_A
\eea
but that is okey since the on-shell gauge field strenght is surely subject to the constraint
\bea
\bar{F}_{\mu\nu} &=& \t {\bar{F}_{\mu\nu}}
\eea
so we find closure on the $\N=5$ equation of motion for the gauge field.

\subsubsection{Commuting $\N=1$ with $\N=5$} 
Let us next commute one $\N=5$ with one $\N=1$ variation. Here we get only the gauge variation term, up to equations of motion. We begin with the scalar fields. Using (\ref{monopole}), we get
\bea
[\delta_{\eta}^1, \Delta_{\epsilon}^5] Z^A &=& Z^A \bar{\Lambda}
\eea
with 
\bea
\bar{\Lambda} &=& -2i \bar{\epsilon}^m \eta (\Sigma^{m6})^B{}_C \t Z^C Z_B,
\eea
but by using the constraint
\bea
Z^A \t Z^C Z_B &=& Z^A Z^C \t Z_B
\eea
we may just as well consider gauge parameter
\bea
\bar{\Lambda} &=& -i \bar{\epsilon}^m \eta (\Sigma^{m6})^B{}_C \(\t Z^C Z_B + Z^C \t Z_B\)
\eea
with the tildes being symmetrically distributed on the two terms.

This is motivated when we turn to the closure on the gauge field. We get
\bea
[\delta_{\eta}^1, \Delta_{\epsilon}^5] \bar{A}_{\mu} &=& - D_{\mu} \bar{\Lambda}
\eea
with gauge parameter
\bea
\bar{\Lambda} &=& i \bar{\epsilon}^m \eta (\Sigma^{m6})_A{}^B \(\t Z^A Z_B + Z^A \t Z_B\)
\eea
with the tildes being symmetrically distributed on two terms. 

We finally turn to the fermions. We have the freedom to place the tildes at our wish in the variation of the fermion since we have already shown that all the different variants are all the same due to constraints. The form that will be most convenient here is the one where the tilde sits on the fermion, or by acting by a monopole again,   
\bea
\delta_{\eta}^1 \psi_A &=& \gamma^{\mu} \eta_{AB} \t {D_{\mu} Z^B} - \eta_{AB} \t {Z^B Z^C Z_C} - \eta_{BC} \t {Z^B Z^C Z_A}
\eea
With this it follows by using the constraints
\bea
\t \psi_A \t Z^C \t Z_B &=& \psi_A \t Z^C Z_B,\cr
\t \psi_A Z^C Z_B &=& \psi_A Z^C \t Z_B
\eea
that we find a gauge variation term with tildes being symmetrically distributed. Moreover by using the constraints
\bea
Z^B \psi_C \t Z_D &=& Z^B \t \psi_C Z_D,\cr
Z^B \t Z C \psi^D &=& Z^B Z^C \t \psi^D,\cr
\t Z^B \psi_C Z_D &=& Z^B \psi_C Z_D,\cr
\t Z^B Z^C \psi^D &=& Z^B Z^C \t \psi^D,\cr
\t Z^B \psi_{[D} Z_{C]} &=&  Z^B \t \psi_D Z_C,\cr
\t Z^B Z^C \psi^D &=& Z^B Z^C \t \psi^D
\eea
we can bring the tilde to the fermion in all the terms. They all follow by either using $X Z^A Y = \t X Z^A \t Y$ or $Z^A X Y = Z^A \t X \t Y$ except for the fifth constraint which can be shown by the following steps,
\bea
\t Z^B \psi_{[D} Z_{C]} = -\t Z^B \t Z_{[C} \t \psi_{D]} = - Z^B \t Z_{[C} \psi_{D]} = Z^B \t \psi_{[D} Z_{C]}
\eea
Since we can bring the tilde to the fermion in all terms, it follows that we have closure on the equation of motion
\bea
\gamma^{\mu} D_{\mu} \t \psi_A + \t \psi_A Z^C Z_C - 2 \t \psi_C Z^C Z_A + \epsilon_{ABCD} Z^B Z^C \t \psi^D &=& 0.
\eea
By applying an overall monopole field, we get
\bea
\gamma^{\mu} D_{\mu} \psi_A + \psi_A Z^C Z_C - 2 \psi_C \t Z^C \t Z_A + \epsilon_{ABCD} \t Z^B \t Z^C \psi^D &=& 0
\eea
where we also used the constraint $Z^C Z_C = \t Z^C \t Z_C$. We can bring this into the equation of $\N=5$ supersymmetric theory by using constraints
\bea
\t \psi_C \t Z^C Z_A = -\t Z_A \t Z^C \psi_C = Z_A \t Z^C \t \psi_C = Z_A Z^C \psi_C,\cr
\t Z^C \t Z_C \psi_A = Z^C \t Z_C \t \psi_A = Z^C Z_C \psi_A
\eea
This finishes the demonstration that we get supersymmetry enhancement to $\N=6$ if we have a monopole with the above mentioned properties which changes the reality condition into the opposite one. 

Since we thus obtain a not manifestly $\N=6$ supersymmetric theory using monopole fields, one could expect this to also have a dual formulation in terms of a manifestly $\N=6$ supersymmetric ABJM theory \cite{Aharony:2008gk}. It would be interesting to understand how the gauge group and the Chern-Simons level changes when we go from the original the $\N=5$ theory to the dual $\N=6$ ABJM theory.

\vskip 2truecm

\subsection*{Acknowledgements}
I would like to thank Soo-Jong Rey, Seok Kim, Sangmin Lee, Jeong-Hyuck Park for discussions. This work was supported by the National Research Foundation of Korea(NRF) grant funded by the Korea government(MEST) through the Center for Quantum Spacetime(CQUeST) of Sogang University with grant number 2005-0049409.

\newpage

\newpage
\section{Appendix}
\subsection{Hermitian three-algebra structure constants}
To understand the connection between three-algebra and its associated Lie algebra, we define
\bea
t^a{}_b(\bullet) &=& [\bullet,T^a;T^b].
\eea
We find that 
\bea
t^a{}_b(T^c) &=& f^{ca}{}_{bd}T^d.
\eea
The hermitian fundemental identity can now be expressed as
\bea
[t^a{}_b,t^c{}_d] &=& f^{ca}{}_{be}t^e{}_d-f^{ae}{}_{db}t^c{}_e
\eea
which means that we have closure among $t^a{}_b$ under the Lie bracket. The multiplication involved in the Lie bracket is that of composition of maps, which is always associative. Consequently the commutator will satisfy the Jacobi identity and so $t^a{}_b$ generate a Lie algebra. It could be that the $t^a{}_b$ is a too big set of generators, for instance most of these may not be anti-hermitian. Let therefore $\Gamma_A$ denote a projector on a linearly independent set of anti-hermitian generators 
\bea
t_A &=& (\Gamma_A)^a{}_b t^b{}_a
\eea
and let $\Sigma^A$ be the inverse of $\Gamma_A$ on the anti-hermitan subset of all the $t^a{}_b$'s. Such anti-hermitian $t^a{}_b$ are then expressible as
\bea
t^a{}_b &=& (\Sigma^A)^a{}_b t_A
\eea
Let us present the Lie algebra as
\bea
[t_A,t_B] &=& C_{AB}{}^C t_C.
\eea
We rise and lower adjoint index $A$ by the Killing form 
\bea
g_{AB} &=& \tr_{adj}(t_A t_B)
\eea
and by its inverse. Let us introduce one last concept, which is the fundamental representation of the Lie algebra. We define a matrix $(t_A)^a{}_b$ associated with $t_A$ as
\bea
t_A(T^a) &=& -(t_A)^a{}_b T^b
\eea
Then it follows from 
\bea
t_A t_B (T^a) &=& (t_B)^a{}_b (t_A)^b{}_c T^c
\eea
that 
\bea
(t_A)^b{}_a
\eea
is a representation of the Lie algebra. The minus sign is because the ordering of multiplication gets reversed, giving an extra minus sign in the Lie bracket. 

We have shown that the fundamental identity implies there is an associated Lie algebra. Does any Lie algebra correspond to a hermitian three-algebra then? As the name indicates that is not the case. However if we restrict ourselves to Lie algebras with anti-hermitian generators, then this is indeed true. We will now proceed to construct the hermitian three-algebra structure constants out of the generators of any hermitian Lie algebra.

We contract the fundemental identity by $(\Gamma_A)^b{}_a$,
\bea
[t_A,t^c{}_d] &=& -(t_A)^c{}_g t^g{}_d + (t_A)^g{}_d t^c{}_g
\eea
We then substitute $t^c{}_d = t_B (\Sigma^B)^c{}_d$,
\bea
[t_A,t_B] (\Sigma^B)^c{}_d &=& \(-(t_A)^c{}_g (\Sigma^B)^g{}_d + (t_A)^g{}_d (\Sigma^B)^c{}_g\) t_B
\eea
We can write this more compactly as
\bea
[t_A,\Sigma^C] &=& -C_{AB}{}^C \Sigma^B.
\eea
This is the fundamental identity that we have just written in a different form. 
The three-algebra structure constants are obtained as
\bea
f^{bc}{}_{da} &=& -(\Sigma^A)^d{}_c (t_A)^b{}_a
\eea
which follows by expanding $t^a{}_b(T^c) = f^{ca}{}_{bd}T^d$.

We solve the fundamental identity by making the most general ansatz 
\bea
\Sigma^A &=& X^{AB} t_B
\eea
Why this is the most general ansatz is not entirely obvious. We discuss this point in more detail when we do the same thing for quaternionic three-algebra below. (We note that $\Sigma_A = g_{AB} \Sigma^B$ close among themselves upon commutation with any Lie algebra element, suggesting these are elements of the Lie algebra.) Plugging this ansatz back into the second equation gives
\bea
[X,C_A] &=& 0.
\eea
By Schur's lemma we then have
\bea
X^{AB} &=& \lambda_l g^{AB}_l.
\eea
with one numerical constant $\lambda_l$ for each simple Lie algebra factor Lie($G_l$) where the Killing form is given by $g_{AB}^l$. The result is
\bea
f^{bc}{}_{da} &=& \sum_l \lambda_l g^{AB}_l (t_A)^b{}_a (t_B)^c{}_d
\eea
This satisfies the fundamental identity for constants $\lambda_l$ and any Lie algebra $G = G_1 \times \cdots \times G_L$ which may include $U(1)$ factors.

The ABJM theories also require that
\bea 
f^{bc}{}_{da} &=& -f^{cb}{}_{da}
\eea
This constraint gives the classification of ABJM theories. If we drop this constraint, then any Lie algebra gives a hermitian three-algebra (where antisymmetry of the h-bracket is thus dropped in the definition of hermitian three-algebra).

\subsection{Symplectic three-algebra structure constants}
We now repeat step by step what we did for hermitian three-algebra in order to solve the quaternionic fundamental identity. We define
\bea
t^{ab}(\bullet) &=& \{\bullet,T^a,T^b\}.
\eea
The fundamental identity can be written as
\bea
[t^{ef},t^{bc}](T^a) &=& f^{bef}{}_g t^{gc}(T^a) - f^{cef}{}_g t^{bg}(T^a)
\eea
showing that $t^{ab}$ generate a Lie algebra. Not all $t^{ab}$ must be linearly independent. Let $\Gamma^A_{ab}$ be a projector on the linearly independent generators
\bea
t^A &=& \Gamma^A_{ab} t^{ab}.
\eea
These generate a Lie algebra 
\bea
[t^A,t^B] &=& C^{AB}{}_C t^C.
\eea
Let us also pick some matrix $\Sigma_A^{ab}$ such that
\bea
t^{ab} &=& \Sigma_A^{ab} t^A
\eea
If $t^{ab}$ as a set contains multiple copies of elements, then there are many equivalent choices of $\Sigma_A^{ab}$. The set $t^A$ on the other hand, does not contain multiple copies of any element, as these elements are linearly independent. Let us define a representation of the Lie algebra through the relation
\bea
t^A(T^a) &=& t^{Aa}{}_b T^b.
\eea
We then note that 
\bea
t^A (t^B (T^a)) &=& t^{Ba}{}_b t^{Ab}{}_c T^c
\eea
so that the ordering of multiplication in this representation becomes opposite to the ordering in which the abstract generators act. We can now express the fundamental identity in the following equivalent form,
\bea
C^{AB}{}_C \Sigma_B^{ab} &=& (t^A \Sigma_C)^{ab} + (t^A \Sigma_C)^{ba},\cr
[t^A,t^B] &=& C^{AB}{}_C,\cr
f^{cab}{}_d &=& \Sigma_A^{ab} t^{Ac}{}_d.
\eea
Let us now assume that 
\bea
t^{ab} &=& t^{ba}
\eea
We then have 
\bea
(\Sigma^A\Sigma^B)^{ab} + (\Sigma^A\Sigma^B)^{ba} &=& [\Sigma^A,\Sigma^B]^{ab}
\eea
and we find 
\ben
[t^A,\Sigma^B] &=& C^{AB}{}_C \Sigma^C.\label{second}
\een
Let us choose Cartan-Weyl basis for the Lie algebra generated by $t^A = (h^i,e^{\alpha})$ where
\bea
[h^i,h^j] &=& 0,\cr
[h^i,e^{\alpha}] &=& \alpha^i e^{\alpha}.
\eea
Let us write $\Sigma^A = (h^i_{\Sigma},e^{\alpha}_{\Sigma})$. Then since the same structure constants appear in (\ref{second})
\bea
[h^i,e^{\alpha}_{\Sigma}] &=& \alpha^i e^{\alpha}_{\Sigma}
\eea
we conclude that
\bea
e^{\alpha}_{\Sigma} &=& e^{\alpha} + Cartan
\eea
Moreover 
\bea
[h^i,h^j_{\Sigma}] &=& 0
\eea
and we conclude that 
\bea
h^j_{\Sigma} &=& X^i{}_j h^j + k^i
\eea
where $k^i$ can be commuting elements in a Cartan subalgebra of some larger Lie algebra whereof $t^A$ generate a subalgebra. We exclude this possibilty by considering the last commutation relation
\bea
[e^{\alpha},e^{-\alpha}] &=& \alpha^i h^i
\eea
since $e_{\Sigma}^{\alpha} = e^{\alpha}+Cartan$ and hence $h^i_{\Sigma} = h^i$. We conclude that the most general solution is to take 
\bea
\Sigma^A &=& \lambda t^B
\eea
where $\lambda$ is a numerical constant, one for each simple Lie algebra. This can be verified also by making the ansatz
\bea
\Sigma^A &=& X^A{}_B t^B
\eea
in which case (\ref{second}) yields the condition
\bea
[X,t^A] &=& 0
\eea
showing that $X$ must be proportional to unit element in each simple Lie algebra by Schur's lemma. The most general form of the three-algebra structure constants is therefore on the form
\bea
f^{abcd} &=& g_{AB} (t^A)^{bc} (t^B)^{ad}
\eea
where $t^A$ generate a Lie algebra and $g_{AB}$ denotes a Killing form on that Lie algebra.

\subsection{Something on $Sp(N)$}\label{sp1}
We define $Sp(N)$ as having dimension $N(2N+1)$. In particular then $Sp(1)$ has dimension $3$ and is the same as $SU(2)$. We have the antisymmetric invariant tensor $J_{ab}$ and its inverse ${(J^{-1})}^{ab}$ in the defining representation where indices ranges as $a = 1,...,2N$. Invariance of $J_{ab}$ means that the $Sp(N)$ generators are defined by the constraint
\bea
(t_A)^c{}_a J_{cb} + (t_A)^c{}_b J_{ac} &=& 0.
\eea
We define 
\bea
G_{AB} &=& (t_A)^b{}_a (t_B)^a{}_b
\eea
and denote its inverse by $G^{AB}$. We must have that
\bea
G^{AB} (t_A)^b{}_a (t_B)^c{}_d &=& A {(J^{-1})}^{bc} J_{ad} + B \delta^b_a \delta^c_d + C \delta^b_d \delta^c_a
\eea
for certain constants $A$, $B$ and $C$. This is the most general invariant tensor we can write down. To determine these coeffients we first note that
\bea
(t_A)_a{}^b &=& J_{ac} (t_A)^c{}_d {(J^{-1})}^{db} 
\eea
generate the same Lie algebra, that is with the same structure constants, as $(t_A)^a{}_b$. This in particular means that
\bea
G_{AB} &=& (t_A)_b{}^a (t_B)_a{}^b
\eea
Then we can contract indices or use $J$ to contract indices in three different way in our ansatz, leading to three equations. These equations can be solved uniquely and gives the result
\bea
G^{AB} (t_A)^b{}_a (t_B)^c{}_d &=& \frac{1}{2} \({(J^{-1})}^{bc} J_{ad} + \delta^b_d \delta^c_a\).
\eea

\subsection{The h-bracket}\label{halge}
We here check that 
\bea
[x,y;z] &=& \alpha x z^{\dag} y + \beta y z^{\dag} x
\eea
satisfies the hermitian fundamental identity 
\bea
[[x,y;z],u;v] &=& [[x,u;v],y;z] + [x,[y,u;v];z] - [x,y;[z,v;u]]
\eea
for any choice of parameters $\alpha$ and $\beta$. First we check that this is a solution for $\alpha=0$,
\bea
uv^{\dag}yz^{\dag}x &=& yz^{\dag}uv^{\dag}x + uv^{\dag}yz^{\dag}x - y(vu^{\dag}z)^{\dag}x
\eea
We see the first and the last term in the right hand side cancel. We next assume that $\alpha\neq 0$. We can then rescale and consider the h-bracket
\bea
[x,y;z] &=&  x z^{\dag} y + \alpha y z^{\dag} x
\eea
We then compute the left hand side
\bea
xz^{\dag}yv^{\dag}u + \alpha(yz^{\dag}xv^{\dag}u + uv^{\dag}xz^{\dag}y) + \alpha^2 uv^{\dag}yz^{\dag}x
\eea
and then each of the terms in the right hand side
\bea
xv^{\dag}uz^{\dag}y + \alpha (uv^{\dag}xz^{\dag}y + yz^{\dag}xv^{\dag}u) + \alpha^2 yz^{\dag} u v^{\dag} x,\cr
xz^{\dag}yv^{\dag}u + \alpha (xz^{\dag}uv^{\dag}y + yv^{\dag}uz^{\dag} x) + \alpha^2 uv^{\dag}yz^{\dag} x,\cr
-xv^{\dag}uz^{\dag}y - \alpha (xz^{\dag}uv^{\dag}y + yv^{\dag} u z^{\dag} y) - \alpha^2 yz^{\dag}uv^{\dag}x
\eea
We then see that the sum coincides with the left hand side order by order in $\alpha$ after that many terms have been canceled. 

This shows that this matrix realization of the h-bracket satisfies the hermitian fundamental identity for any choice of parameters $\alpha$ and $\beta$.

\subsection{The q-bracket}\label{qbracket}
We define the q-bracket as
\ben
\{T^{\qa},T^{\qb},T^{\qc}\} &=& [T^{\qa},T^{\qb};\omega T^{\qc}] + [T^{\qa},T^{\qc};\omega T^{\qb}]\label{q}
\een
We assume the hermitian fundamental identity holds for the h-bracket on the right-hand side, and we wish to show that the q-bracket on the left-hand side satisfies the quaternionic fundamental identity. We start by exapnding out
\ben
\{\{T^{\qa},T^{\qb},T^{\qc}\},T^{\qe},T^{\qf}\} &=& [[T^{\qa},T^{\qb};\omega T^{\qc}],T^{\qe};\omega T^{\qf}]\cr
& +& [[T^{\qa},T^{\qc};\omega T^{\qb}],T^{\qe};\omega T^{\qf}]\cr
& +& [[T^{\qa},T^{\qb};\omega T^{\qc}],T^{\qf};\omega T^{\qe}]\cr
& +& [[T^{\qa},T^{\qc};\omega T^{\qb}],T^{\qf};\omega T^{\qe}].\label{out}
\een
We expand each of these four terms using the hermitian fundamental identity. The first of the four terms is expanded as
\bea
[[T^{\qa},T^{\qe};\omega T^{\qf}],T^{\qb};\omega T^{\qc}] + [T^{\qa},[T^{\qb},T^{\qe};\omega T^{\qf}];\omega T^{\qc}] - [T^{\qa},T^{\qb};[\omega T^{\qc},\omega T^{\qf};T^{\qe}]]
\eea
We shall rewrite the last term as
\bea
- [T^{\qa},T^{\qb};[\omega T^{\qc},\omega T^{\qf};T^{\qe}]] &=& [T^{\qa},T^{\qb};\omega [T^{\qc},T^{\qf};\omega T^{\qe}]].
\eea
We then wish to match the result one then gets with what one gets when expanding out
\bea
\{\{T^{\qa},T^{\qe},T^{\qf}\},T^{\qb},T^{\qc}\} + \{T^{\qa},\{T^{\qb},T^{\qe},T^{\qf}\},T^{\qc}\} + \{T^{\qa},T^{\qb},\{T^{\qc},T^{\qe},T^{\qf}\}
\eea
That would then establish that the q-bracket defined by (\ref{q}) satisfies the quaternionic fundamental identity.

We begin expanding out the first term, which is sort of trivial,
\bea
\{\{T^{\qa},T^{\qe},T^{\qf}\},T^{\qb},T^{\qc}\} &=& [[T^{\qa},T^{\qe};\omega T^{\qf}],T^{\qb};\omega T^{\qc}] \cr
&+& [[T^{\qa},T^{\qe};\omega T^{\qf}],T^{\qc};\omega T^{\qb}] \cr
&+& [[T^{\qa},T^{\qf};\omega T^{\qe}],T^{\qb};\omega T^{\qc}] \cr
&+& [[T^{\qa},T^{\qf};\omega T^{\qe}],T^{\qc};\omega T^{\qb}]
\eea
The second term is non-trivial
\bea
\{T^{\qa},\{T^{\qb},T^{\qe},T^{\qf}\},T^{\qc}\} &=& [T^{\qa},[T^{\qb},T^{\qe};\omega T^{\qf}];\omega T^{\qc}] \cr
&+& [T^{\qa},[T^{\qb},T^{\qf};\omega T^{\qe}];\omega T^{\qc}] \cr
&+& [T^{\qa},T^{\qc};\omega[T^{\qb},T^{\qe};\omega T^{\qf}]] \cr
&+& [T^{\qa},T^{\qc};\omega[T^{\qb},T^{\qf};\omega T^{\qe}]]
\eea
and likewise the last term
\bea
\{T^{\qa},T^{\qb},\{T^{\qc},T^{\qe},T^{\qf}\}\} &=& [T^{\qa},T^{\qb};\omega[T^{\qc},T^{\qe};\omega T^{\qf}]]\cr
&+& [T^{\qa},T^{\qb};\omega [T^{\qc},T^{\qf};\omega T^{\qe}]]\cr
&+& [T^{\qa},[T^{\qc},T^{\qe};\omega T^{\qf}];\omega T^{\qb}]\cr
&+& [T^{\qa},[T^{\qc},T^{\qf};\omega T^{\qe}];\omega T^{\qb}].
\eea
As it turns out, each of these $12$ terms can be matched with a corresponding term we obtain by expanding out (\ref{out}).

\end{document}